\documentclass[twocolumn, trackchanges]{aastex7}

\usepackage{amsmath}
\begin{document}

\title{Circular Mystery: Exploring Diffuse Emission Surrounding a Radio Galaxy with uGMRT and VLA Multiwavelength Observations}

\shorttitle{Circular Diffuse Source}
\shortauthors{Kumari et al.}

\author[orcid=0000-0003-4213-9679]{Shobha Kumari}
\affiliation{Midnapore City College, Kuturia, Bhadutala, Paschim Medinipur, West Bengal, 721129, India}
\email[show]{shobhakumari@mcconline.org.in} 
\author[orcid=0000-0003-2325-8509]{Sabyasachi Pal}
\affiliation{Midnapore City College, Kuturia, Bhadutala, Paschim Medinipur, West Bengal, 721129, India}
\email[]{sabya.pal@gmail.com} 
\author[orcid=0000-0002-6794-7405]{Souvik Manik}
\affiliation{Midnapore City College, Kuturia, Bhadutala, Paschim Medinipur, West Bengal, 721129, India}
\email[]{souvikmanik@mcconline.org.in}

\begin{abstract}
We report the discovery of J1218+1813, a circular diffuse radio source detected in the Very Large Array (VLA) FIRST survey at 1400 MHz. The source has an angular diameter of approximately one arcminute, corresponding to a physical size of $\sim180$ kpc, and is associated with an elliptical galaxy with a redshift of $z=0.139635$. To investigate its nature, we conducted a comprehensive multiwavelength analysis spanning both high and low radio frequencies, utilizing the VLA in C-configuration at \textit{L}, \textit{C} and \textit{X} bands, along with the upgraded Giant Metrewave Radio Telescope (uGMRT) at bands 3 (250--500 MHz), 4 (550--850 MHz), and 5 (1000--1460 MHz). A spectral study based on these multiwavelength observations reveals that the core of J1218+1813 exhibits a steep spectral index of 0.8, indicating that the emission is dominated by optically thin synchrotron radiation. The spectral index varies between 1.1 and 1.8, from the inner to the outer structure, with the steepest values observed at the periphery of the diffuse emission. An optical analysis of the central host galaxy using spectroscopic data is also performed. The estimated black hole mass is $2.8 \pm 0.8 \times 10^{8}$ \textit{M}${_\odot}$, while the host galaxy has a stellar mass of 2.9 $\times$ 10$^{11}$ \textit{M}${_\odot}$ and a stellar age of 8.58 Gyr. The identification of J1218+1813 is particularly significant because it provides insight into the mechanisms responsible for the formation of circular diffuse radio structures surrounded by an elliptical galaxy. Potential formation scenarios for J1218+1813 are discussed in this paper. 

\end{abstract}


\section{Introduction}
Active galactic nuclei (AGNs) serve as dynamic engines driving the evolution of radio galaxies, profoundly influencing the visibility and behavior of their radio lobes \citep{Kauffmann2003, Hardcastle2020, Morganti2017}.  Understanding the processes that govern AGN activity and its effects on radio-galaxy structures is essential for studying the life cycle of these cosmic phenomena \citep{Best2004, Fabian2012, Hardcastle2018, Perucho2019}. Radio lobes form passively when the jets of radio galaxies stop being active unless they get interrupted by another event in the surrounding medium \citep{Scheuer1974, Kaiser1997, Murgia2011}. Relativistic electrons rapidly lose energy due to inverse Compton scattering and synchrotron radiation, resulting in a fainter radio lobe \citep{Blundell2000, Mocz2011, Turner2018}. During periods of around a few million years, the lobe may become invisible.

The interplay between AGN activity and the surrounding environment is crucial in shaping the evolutionary pathways of radio galaxies. AGN jets inject vast amounts of energy into the interstellar medium and intergalactic medium (IGM) of their host galaxy, influencing star formation, gas dynamics, and large-scale structure formation \citep{McNamara2007, Fabian2012, Gaspari2020}. These energetic outflows create cavities, shock fronts, and turbulence in the hot gas environment, which can suppress cooling flows and regulate galaxy growth \citep{Croston2005, Hardcastle2013, Gitti2012}. Furthermore, the duration and recurrence of AGN activity play a pivotal role in the observed morphological diversity of radio galaxies, ranging from Fanaroff-Riley (FR) classes to more complex hybrid structures such as hybrid
morphology radio sources (HyMoRS) and restarted radio galaxies \citep{GopalKrishna2000, Saikia2009, Sh22}. In cases where AGN activity ceases, the aging radio lobes experience spectral steepening as relativistic particles lose energy, ultimately transitioning into faint, diffuse emission structures known as ``fossil" or ``relic" radio sources \citep{Murgia2011, Shulevski2017}.

Recent discoveries of peculiar circularly symmetric structures known as odd radio circles (ORCs) \citep{No21b, No21c, Ko21, Om22c, Lo23, Ku23a, Ku23b, Ko24a, Ko24b, No25} have sparked interest in their possible connection to AGN evolutionary processes. These enigmatic sources may represent a previously overlooked stage in the life cycle of radio galaxies or may arise from distinct physical mechanisms. Several hypotheses have been proposed to explain ORC formation, including shocks driven by AGN outflows \citep{Fu24}, galaxy mergers producing large-scale circular features \citep{Do23}, and shock acceleration in remnant radio lobes forming re-energized vortex rings (phoenixes) \citep{Sha24}. Other proposed mechanisms include virial shocks and AGN jet-inflated bubbles \citep{Ya24, Li24}, further linking ORCs to AGN-driven dynamics and large-scale structure formation. Multiwavelength observations remain crucial in distinguishing these scenarios and understanding the physical processes that shape ORC morphology.

\begin{figure}
\vbox{
\centerline{
\includegraphics[width=9.0cm, origin=c]{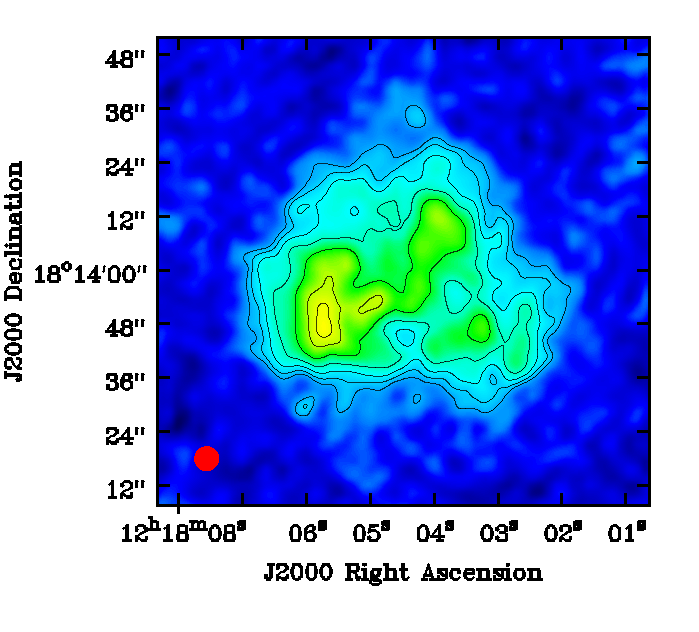}
}
}
\caption{VLA FIRST image of J1218+1813 at 1400 MHz. The contour levels are at 3$\sigma$ $\times$(1, 1.4, 2, 2.8, 4, 4.5, 5, 5.4, 5.7, 5.8, 5.9, 6, 6.2, 6.5, 7, 7.2, 7.4), where $\sigma$ = 140 $\mu$Jy beam$^{-1}$ is the rms noise near the background region of J1218+1813. In the left corner of the image, the synthesized beam with a resolution of 5.4$'' \times 5.4''$ is shown.}
\label{fig:VLA_FIRST}
\end{figure}

In the present article, we report the identification of a circular diffuse radio source, J1218+1813, in the Very Large
Array (VLA) FIRST survey at 1400 MHz \citep{Be95, Wh97}. Figure \ref{fig:VLA_FIRST} displays the 1400 MHz VLA FIRST image of J1218+1813, revealing its circular morphology with an angular diameter of 70$''$, corresponding to a physical size of $\sim180$ kpc. 
This diffuse source is associated with a red elliptical galaxy, SDSS J121804.88+181353.7 (also cataloged as 2MASX J12180486+1813533, PGC1553747, and SDSS J121804.89+181353.6). The optical counterpart of J1218+1813 is identified using data from the Panoramic Survey Telescope and Rapid Response System (Pan-STARRS; \citet{Ch16})\footnote{https://ps1images.stsci.edu/cgi-bin/ps1cutouts}, the Sloan Digital Sky Survey (SDSS-DR16) catalog\footnote{https://www.sdss.org/dr16/} \citep{Ahu20}, and the Dark Energy Camera Legacy Survey (DECaLS; \citet{De19}). The host galaxy has a spectroscopic redshift of $z = 0.139635$, obtained from the SDSS.
Additional observations from the TIFR GMRT Sky Survey (TGSS) at 150 MHz \citep{In17} and the National Radio Astronomy Observatory (NRAO) VLA Sky Survey (NVSS) at 1400 MHz \citep{Co98} detected J1218+1813. However, its distinct circular, diffuse morphology was not evident due to the limited sensitivity and low resolution of these surveys. Similarly, while the Very Large Array Sky Survey (VLASS) at 3000 MHz \citep{La20} provides high-resolution imaging, its insufficient sensitivity results in the loss of the extended diffuse structure of J1218+1813.

This paper is organized as follows. Section \ref{sec:data} discusses observation and data reduction. In Section \ref{sec:result}, we elaborate on our results. We discuss our result in Section \ref{sec:discussion}. Section \ref{sec:conclusion} summarizes the current paper. For the basic calculation, we have used the cosmology parameters from the result of final full-mission Planck measurements of the cosmic microwave background anisotropies: $H_0$ = 67.4 km s$^{-1}$ Mpc$^{-1}$, $\Omega_{\text{vac}}$ = 0.685 and $\Omega_m$ = 0.315 \citep{Ag20}.

\begin{figure*}
\vbox{
\centerline{
\includegraphics[width=9.5cm, origin=c]{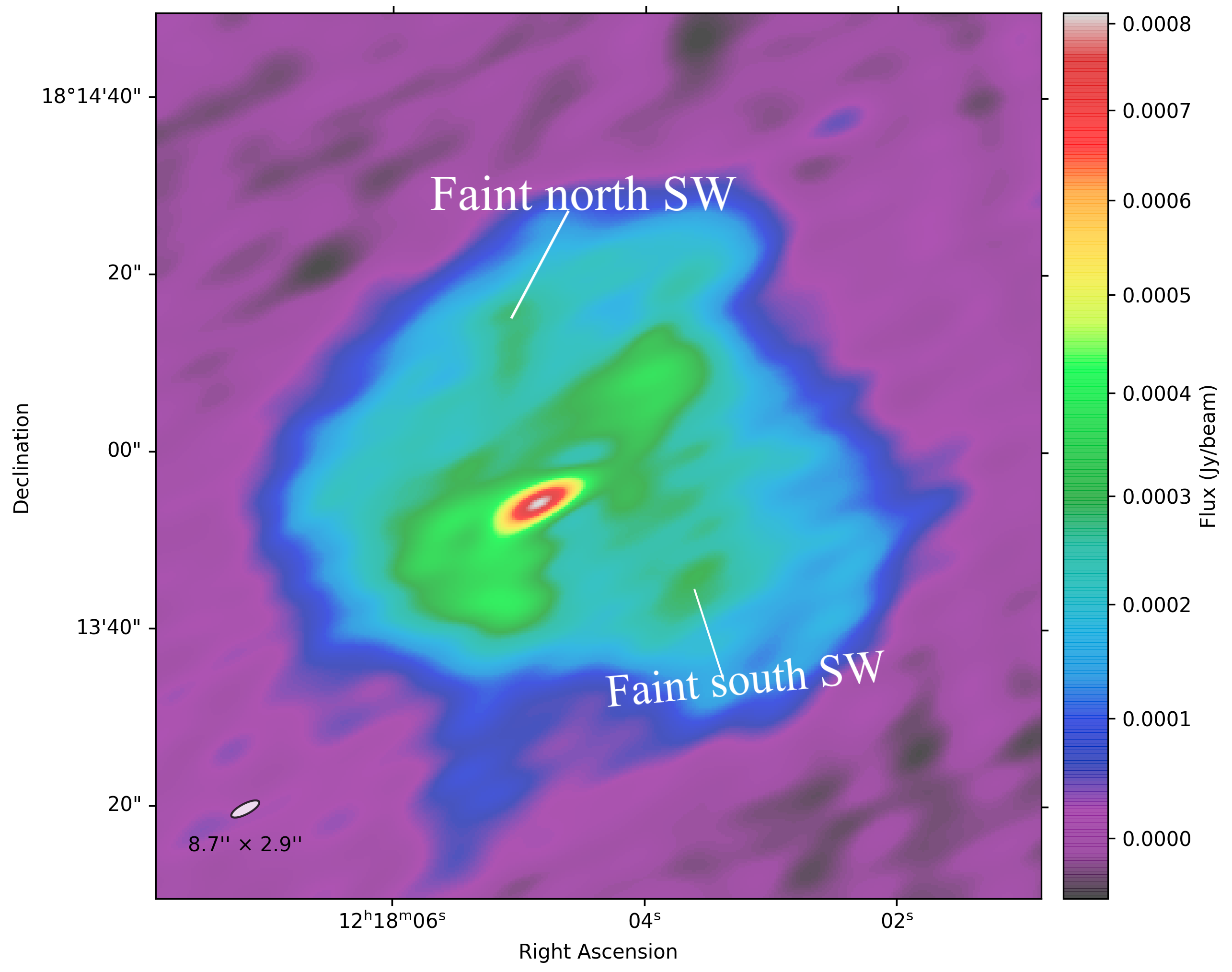}
\includegraphics[width=9.5cm, origin=c]{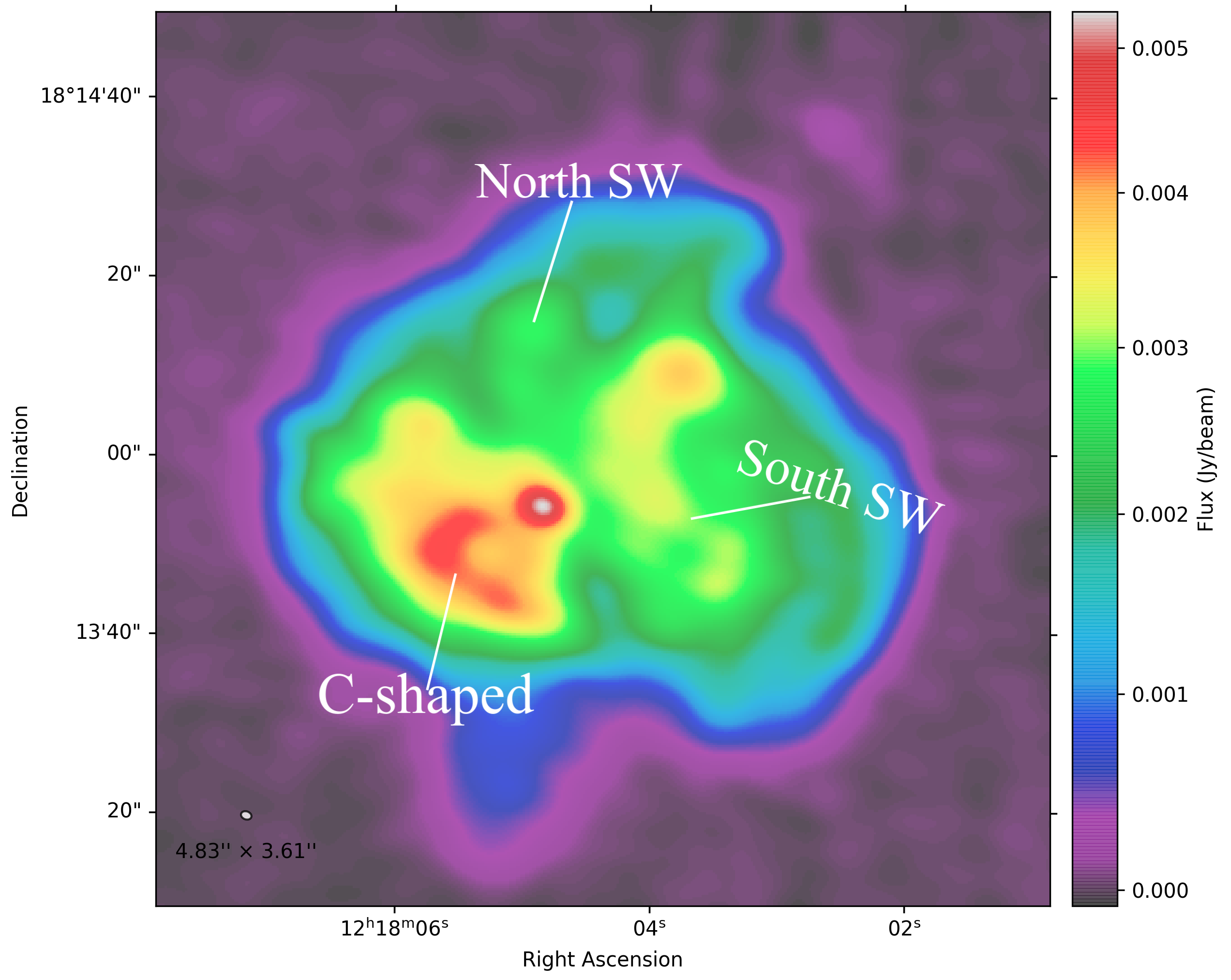}
}
}
\vbox{
\centerline{
\includegraphics[width=9.2cm, origin=c]{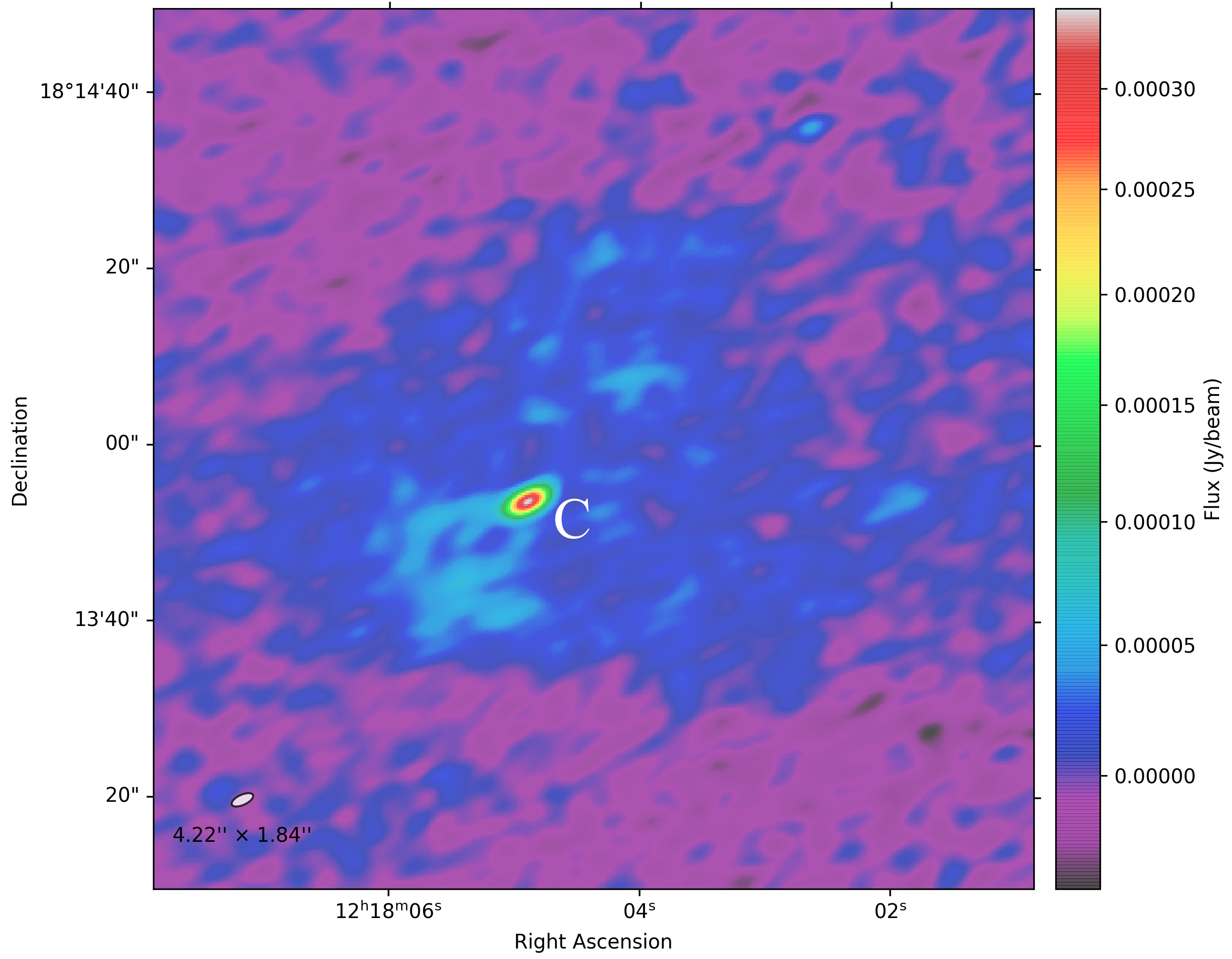}
\includegraphics[width=9.6cm, origin=c]{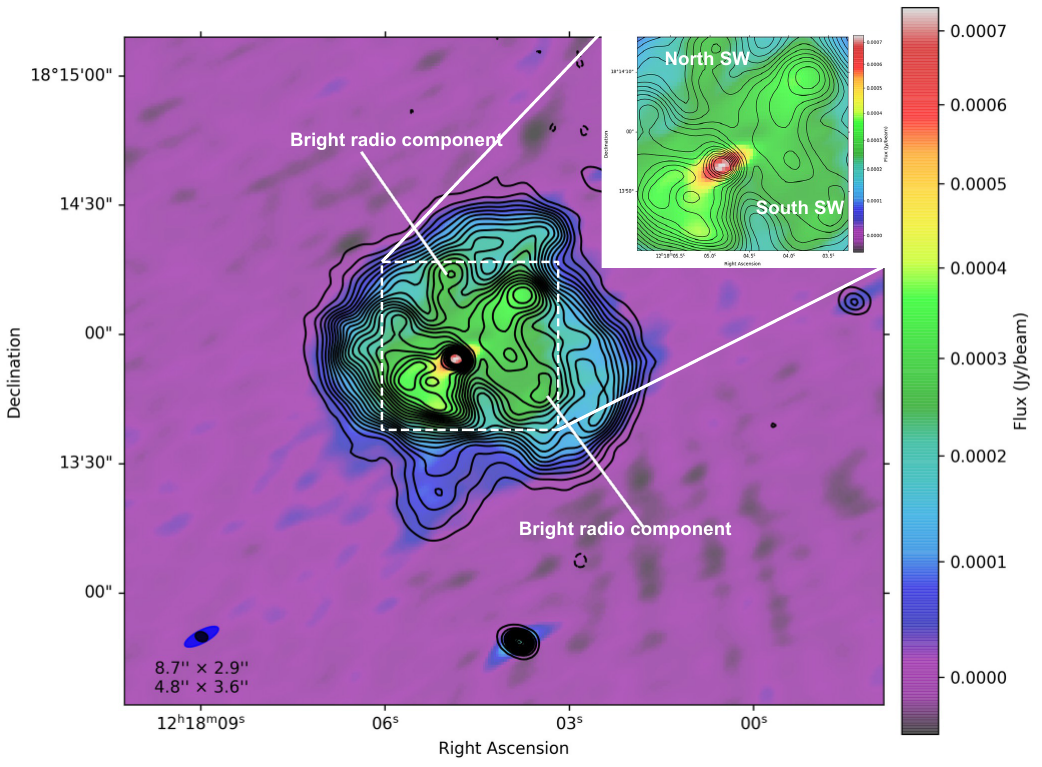}
}
}
\caption{Upper Left: VLA \textit{C} band image of J1218+1813 at 6000 MHz. Upper Right:  GMRT band 4 image of J1218+1813 at 647 MHz. Lower Left: VLA \textit{X} band image of J1218+1813 at 10000 MHz. Lower Right: VLA \textit{C} band image of J1218+1813 at 6000 MHz overlayed with GMRT band 4 image (in contour) at the central frequency 647 MHz. The contour levels are at 3$\sigma\times$(1, 1.4, 2, 2.8, 4, 5.7, 8, 11, 16, 23, 32, 40, 48, 52, 58, 64, 70, 75, 80, 90, 100), where $\sigma$ = 18  $\mu$Jy beam$^{-1}$ is the RMS noise. The color scale on the right side represents the flux density in Jy beam$^{-1}$. The ellipse in the left corner presents the synthesized beam size of the source.}
\label{fig:VLA_GMRT}
\end{figure*}

\begin{figure*}
\vbox{
\centerline{
\includegraphics[width=9.5cm, origin=c]{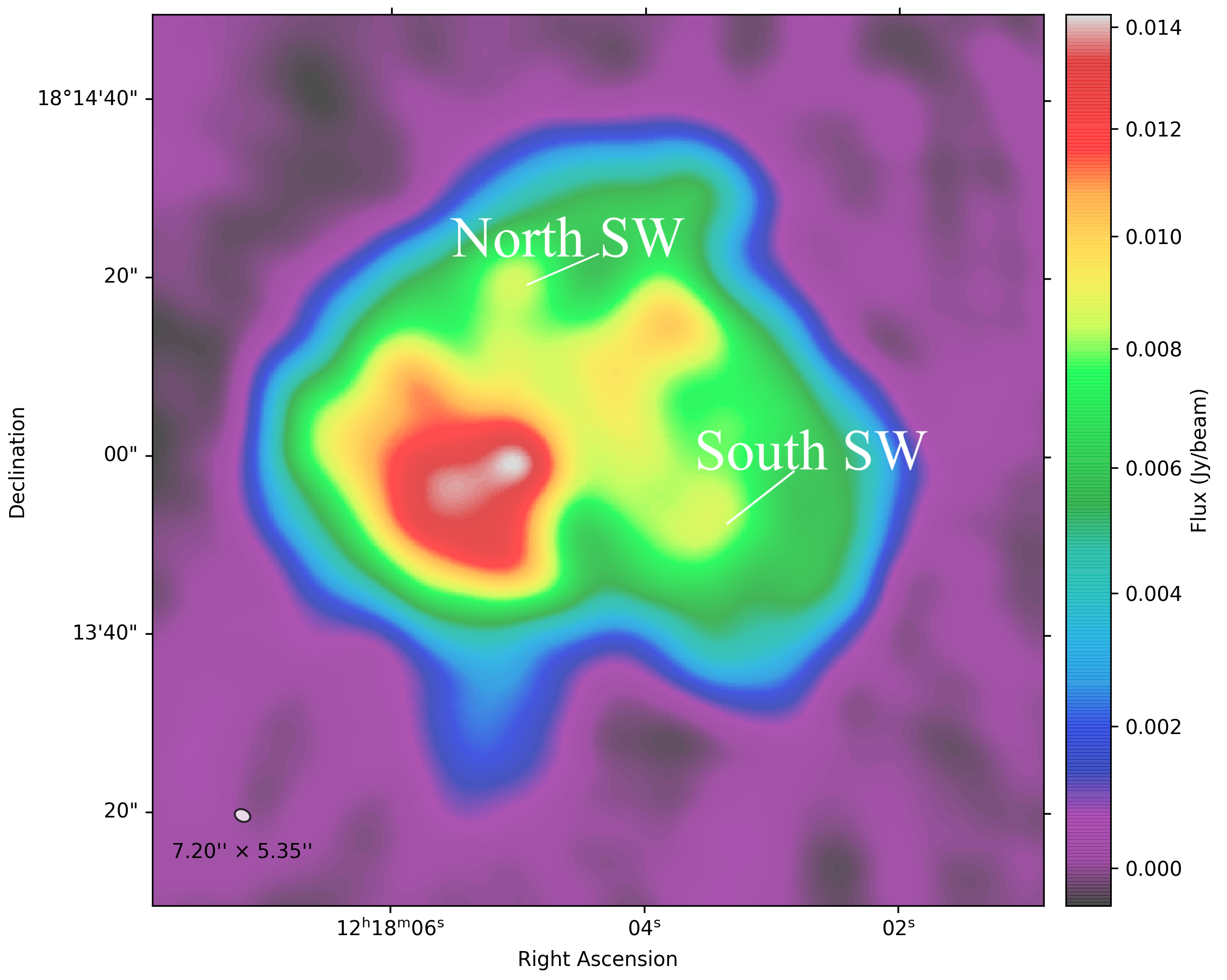}
\includegraphics[width=9.6cm, origin=c]{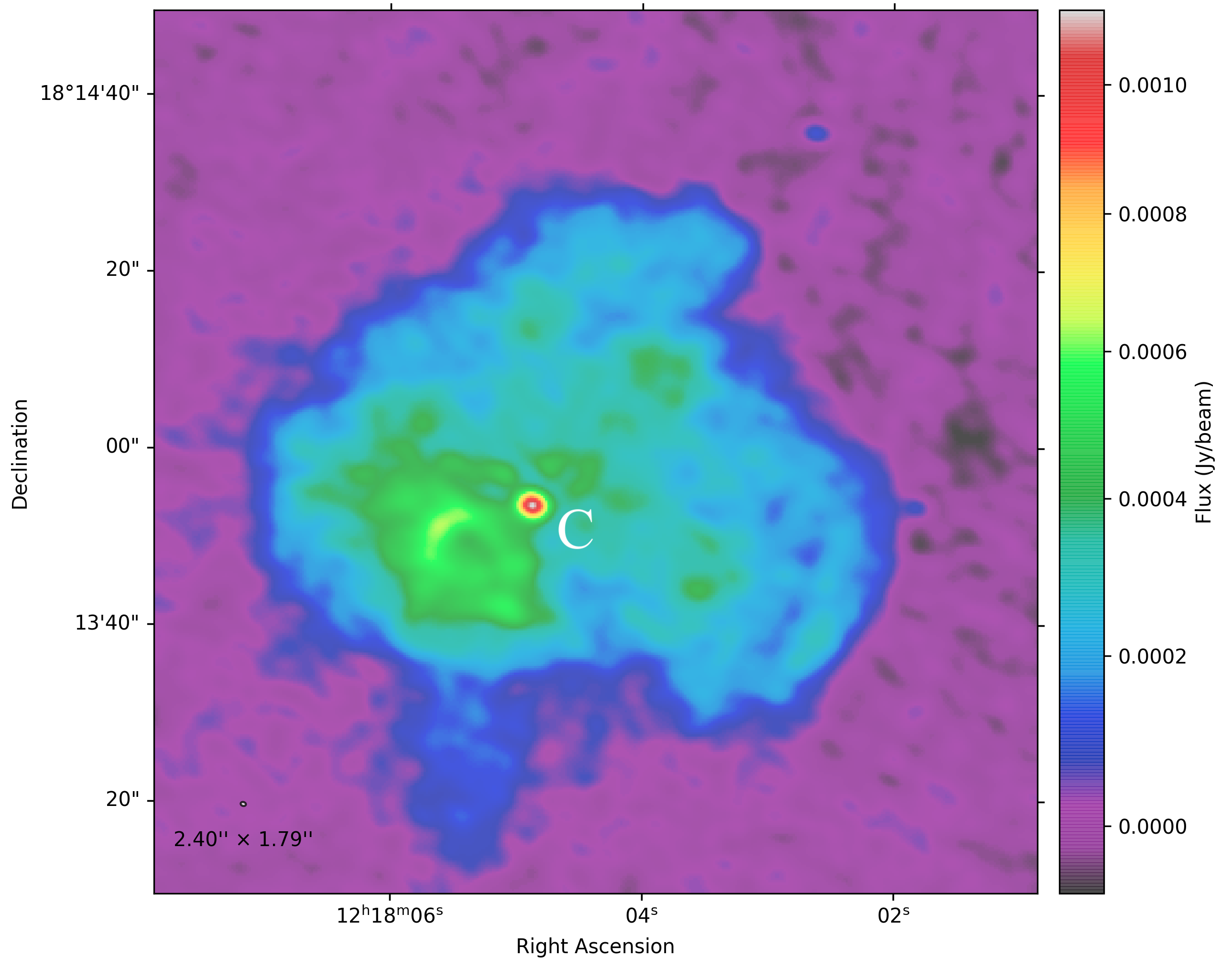}
}
}
\caption{Left: GMRT band 3 image at the central frequency of 402 MHz. Right: GMRT band 5 image at the central frequency of 1274 MHz. The white ellipse in the left corner presents the synthesized beam size of the source.
}
\label{fig:GMRT_band_3_5}
\end{figure*}

\begin{figure*}
\vbox{
\centerline{
\includegraphics[width=18cm, origin=c]{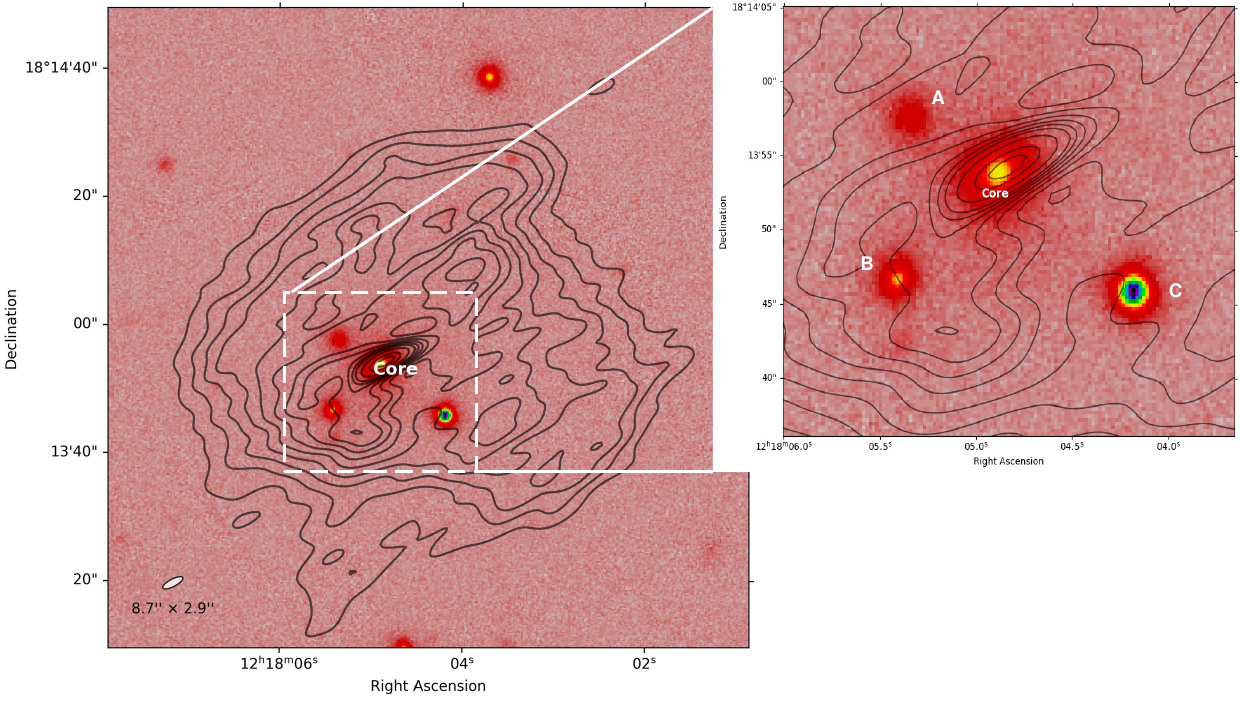}
}
}
\caption{VLA \textit{C} band image of J1218+1813 at 6000 MHz overlayed with an optical image of Dark Energy Spectroscopic Instrument's Legacy Imaging Surveys \cite[DESI LS DR9;][]{Sc21}. The contour levels are at 3$\sigma\times$(1, 1.4, 2, 2.8, 4, 5.7, 8, 11, 16, 23, 26), where $\sigma$ = 11 $\mu$Jy beam$^{-1}$ is the RMS noise. The white ellipse in the left corner presents the synthesized beam size of the source.}
\label{fig:optical}
\end{figure*}

\section{Data and Observations}
\label{sec:data}
This section elaborates on the VLA FIRST survey, wideband follow-up radio observations, and radio data analysis procedures for J1218+1813.
\subsection{VLA FIRST Survey}
\label{sub:FIRST}
The FIRST survey at 1400 MHz (21 cm) spans a vast area of 10,575 square degrees, covering the radio sky near the North and South Galactic Caps. With an angular resolution of 5$''$ and a mean RMS sensitivity of 0.15 mJy \citep{Be95, Wh97}, the survey is well suited for detailed morphological studies of faint radio sources. We utilized the VLA FIRST survey due to its excellent resolution and high sensitivity, which are instrumental in exploring the intricate structures of such sources.

The FIRST database has been extensively used in the past to uncover and study radio galaxies with unique, rare, and irregular morphologies. These include HyMoRS \citep{Sh22}, and winged radio galaxies \citep{Ch07, Ya19, Be20}, head-tailed sources \citep{Mis19, Sa22}, compact steep spectrum sources \citep{Ku02, Ma06}, and giant radio sources \citep{Ku18}. The survey continues to be an invaluable resource for advancing our understanding of the diverse population of radio galaxies in the universe.

\subsection{VLA Observation and Data Analysis}
\label{sub:VLA}
The reported source J1218+1813 in the current paper was observed with the VLA in C-configuration across three frequency bands: \textit{L} band (1000--2000 MHz), \textit{C} band (4000--8000 MHz), and \textit{X} band (8000--12000 MHz) corresponding to the central frequencies of $\sim1.5$ GHz, $\sim6$ GHz and $\sim10$ GHz. Each band observation consisted of two scans lasting 10 and 8 minutes, interspersed with 2--4 minutes of pointing to the phase calibrator (J1125+2610). The flux density scale was set using the calibrator 3C 286, which was observed for 5–-8 minutes during the observing sessions. Data were processed using the Common Astronomy Software
Applications (CASA) \texttt{6.5.0} pipeline, which included calibration steps to ensure data accuracy. Additional manual flagging was applied to remove residual radio frequency interference and noisy scans, enhancing overall data quality for subsequent analysis. We run two rounds of phase-only self-calibration for the source. We imaged the source at 4160 MHz and 6300 MHz using wideband data in the VLA \textit{C} band. The data analysis parameters are tabulated in Table \ref{tab:data}.

\subsection{uGMRT Observation and Data Analysis}
\label{sub:GMRT}
We observed the source using the high-sensitive, low-frequency radio telescope Giant Metrewave Radio Telescope \citep[GMRT;][]{Sw91}, in band 3 (250--500 MHz), band 4 (550--850 MHz), and band 5 (1000--1460 MHz) with approximately 1.5 hr of integration time in each band. The data were recorded with RR and LL correlators in each band with a total observing bandwidth of 200 MHz. Data for each band was collected with 4096 channels. For data analysis, we applied the \texttt{CASA} package \citep{Mc07}. Standard procedures were performed for bandpass calibration, complex gain calibration, and flagging (removal of bad data). For absolute flux calibration, the flux density scale of \citet{Pe17} is used. For better sensitivity, multiple rounds of phase-only analysis and one round of amplitude and phase self-calibration have been performed. Using the Briggs weighting system \citep{Br95}, we imaged the final visibilities with robust = 0. 
The details of the data analysis parameters are summarized in Table \ref{tab:data}. This table includes critical information such as the observing date, central frequency, total bandwidth, on-source integration time, robust parameter, rms noise, half-power beamwidth and position angle. These parameters provide a comprehensive overview of the observational setup and data quality, which helps to interpret and analyze the source properties.

\section{Results}
\label{sec:result}
\begin{table*}
 \caption{Radio Observations and Data Analysis Parameters}
    \centering
    \begin{tabular}{ccccccccc}
    \hline
        Telescope  &Date of Observation&Band&Frequency & Bandwidth&Integration Time&Robust &RMS & HPBW, PA \\
         &(dd/mm/yyyy)& &(MHz)&(MHz)& (min)&&($\mu$Jy beam$^{-1}$)&  ($'' \times '', ^{\circ})$\\
        \hline
        uGMRT &18/05/2024 &3 &~~402 &~200& 75&0 &  40 & 7.2$'' \times 5.4''$, 65.9 \\ 
        uGMRT &19/05/2024 &4&~~647 &~200& 94&0 &  18 & 4.8$'' \times 3.6''$, 70.6 \\
         uGMRT &21/05/2024&5&~1274 & ~200&94 &0 & 15 & 2.4$'' \times 1.8''$, 74.4 \\
         VLA & 22/03/2024&\textit{L}&~1519 &1000 &60&0 & 45 &  40$'' \times 12''$, --61.6 \\
         VLA & 22/03/2024&\textit{C}&~4160 &4000 &60&0 & 35 & 12$'' \times 4.0''$, --63.3 \\
        VLA & 22/03/2024&\textit{C}&~6000 & 4000&60&0 & 11 & 8.7$'' \times 2.9''$, --62.3 \\
         VLA &22/03/2024&\textit{C} &~6300 & 4000&60&0 & 20 & 8.8$'' \times 2.8''$, --63.5 \\
         VLA &22/03/2024 &\textit{X}&10000 & 4000&60&0 & ~8 & 4.2$'' \times 1.8''$, --65.4 \\
         \hline
    \end{tabular}
    \label{tab:data}
\end{table*}

\subsection{Result from VLA Observation}
\label{sec:result_VLA}
The upper left and the lower left panels of Figure \ref{fig:VLA_GMRT} present the map of the VLA \textit{C} band at the central frequency of $\sim 6$ GHz and the VLA \textit{X} band at the central frequency of $\sim 10$ GHz. The angular resolution of 8.7$'' \times 2.9''$ is reached for the final image of the VLA \textit{C} band and the angular resolution of 4.22$'' \times 1.84''$ for the final image of the VLA \textit{X} band. The typical rms values of the final images in the VLA \textit{C} and \textit{X} bands are $\sim$11 $\mu$Jy and $\sim$8 $\mu$Jy, respectively, measured in background regions near the source. The VLA in the \textit{C} band is capable of mapping the inner feature of the source (see the upper left panel of Figure \ref{fig:VLA_GMRT}) with the resolved core. The southeast (SE) side lobe of the inner feature is bent in a C shape, whereas the northwest (NW) side lobe shows no clear structure and displays a random blobiness of the diffuse emission. A very faint secondary wing (SW)-like emission can also be seen in the VLA \textit{C} band image, labeled as Faint North SW and Faint South SW (see the upper left panel of Figure \ref{fig:VLA_GMRT}). The VLA \textit{X} band image effectively identifies the compact radio core, showcasing its high-resolution capabilities to detect small-scale structures. However, it fails to recover the prominent inner structure of J1218+1813 (detected in the VLA \textit{C} band) because the high-frequency observations suffer from flux loss. Due to the low resolution of the image, the circular, diffuse, and inner structure of J1218+1813 was not resolved in the C-configuration VLA L-band image.  

 \begin{figure}
\vbox{
\centerline{
\includegraphics[width=9.0cm, origin=c]{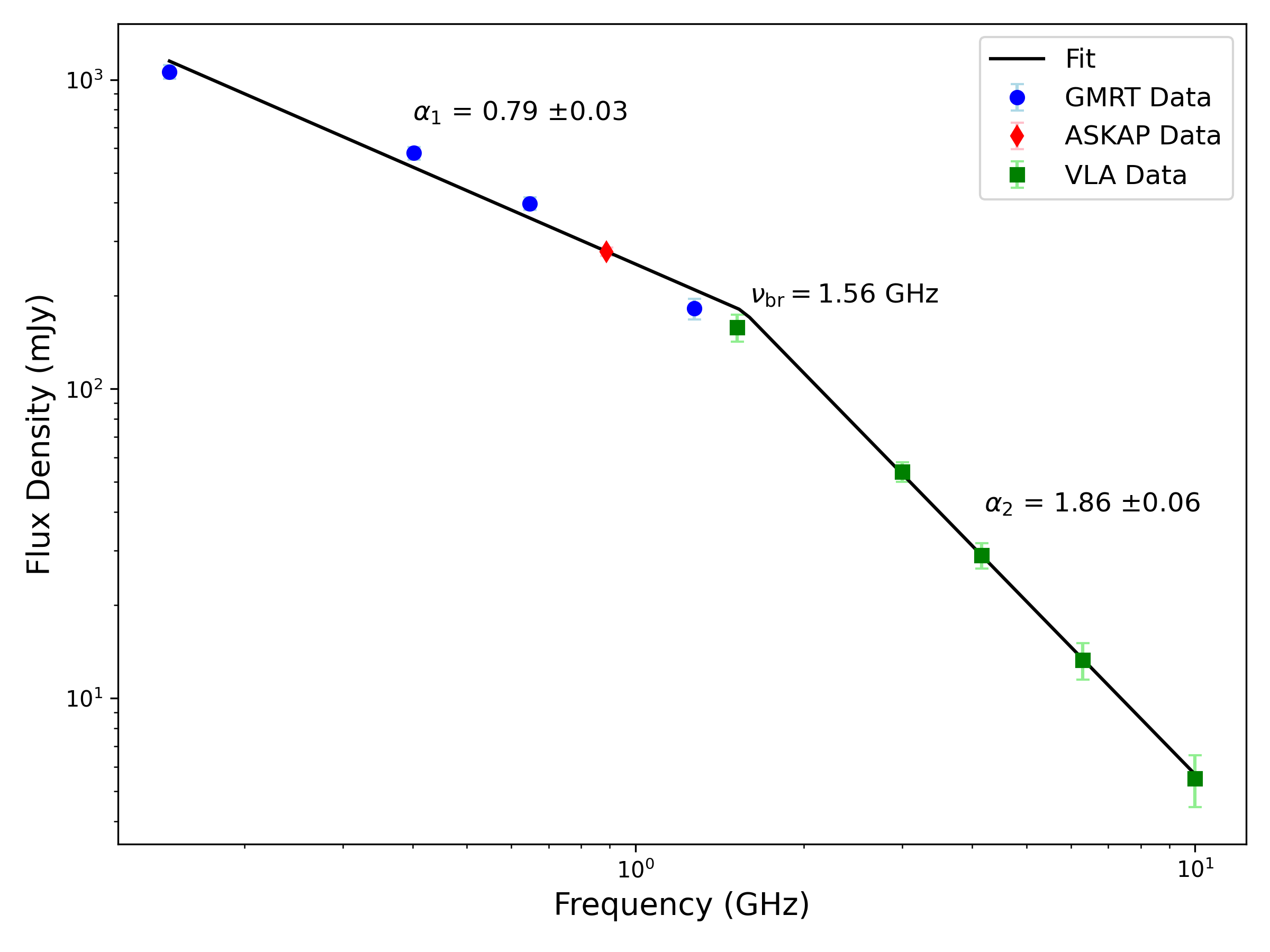}
}
}
   \caption{Integrated radio spectrum for J1218+1813 between 147 MHz to 10,000 MHz.}
\label{fig:spec_map2}
\end{figure}

\begin{figure}
\vbox{
\centerline{
\includegraphics[width=9.0cm, origin=c]{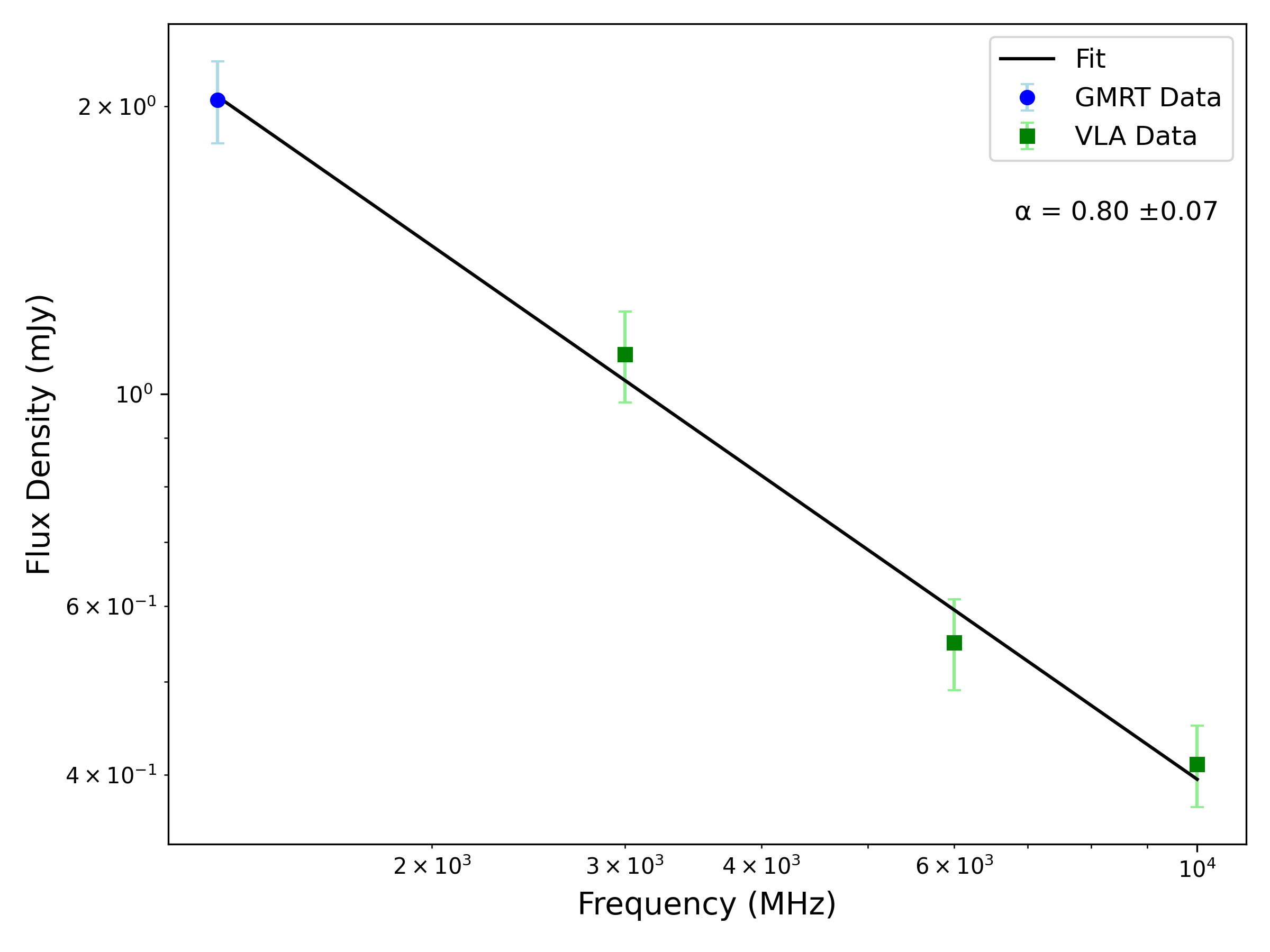}
}
}
\caption{Integrated radio spectrum for radio core of J1218+1813 between 1274 MHz to 10,000 MHz.}
\label{fig:spec_map3}
\end{figure}

\subsection{Result from uGMRT Observation}
\label{sec:result_GMRT}
In the final images of GMRT in bands 3, 4, and 5, we achieved angular resolutions of 7.2$'' \times 5.4''$, 4.8$'' \times 3.6''$, and 2.4$'' \times 1.8''$, respectively. The typical rms noise levels measured in the background regions close to the source are approximately 40 $\mu$Jy, 18 $\mu$Jy, and 15 $\mu$Jy for bands 3, 4, and 5, respectively. The GMRT band 4 image, shown in the upper right panel of Figure \ref{fig:VLA_GMRT}, offers a detailed view of the structure of J1218+1813. It reveals features similar to the VLA \textit{C} band image, including the inner structure. The GMRT band 4 image confirms the wing-like secondary emission along the minor axis of the source, which was seen very faintly in the VLA \textit{C} band image. At the edges (north and south) of the SW, a bright radio component can also be seen (see the lower right panel of Figure \ref{fig:VLA_GMRT}). The GMRT image at band 3 also confirms this wing-like emission in the source (displayed in the left panel of Figure \ref{fig:GMRT_band_3_5}). The GMRT band 5 image missed the inner structure and wing-like emission of the source due to loss in flux, similar to the VLA X-band image (displayed in the right panel of Figure \ref{fig:GMRT_band_3_5}).

\subsection{Optical Counterpart}
\label{subsec:optical}
Figure \ref{fig:optical} represents the superimposed image of J1218+1813 between radio (taken VLA \textit{C} band image) and optical (taken from the Pan-STARRS). In this superimposed image, the optical core is denoted by ``Core'' (see Figure \ref{fig:optical}). As can be seen in the zoomed panel of Figure \ref{fig:optical}, the optical host galaxy (denoted by ``Core'') coincides with the radio core. The optical galaxy has a spectroscopic redshift of $z=0.139635$. Three nearby optical galaxies (denoted by ``A", ``B", and ``C" in Figure \ref{fig:optical}) can be seen in the image. The optical galaxies ``A", ``B", and ``C" have photometric redshifts of 0.183$\pm0.036$, 0.151$\pm0.015$, and 0.213$\pm0.029$, respectively. The photometric redshifts are taken from DECaLS data release 9 \citep[DR9;][]{De19}. Considering the photometric redshifts of A, B, and C optical galaxies, the relative velocities ($v=c\delta z/(1+z)$) of these galaxies are greater than 2000 km/s with respect to the central host galaxy. This suggests that these galaxies are possibly not interacting or associated with the host galaxy unless they are members of any rich galaxy cluster \citep{Fe21}, but we need spectroscopic redshift of A, B, and C galaxies to confirm this statement. 
 
\subsection{Search for Cluster or Group Environments}
\label{sub:Cluster}
We used NED to search for available galaxy clusters within 10$'$ of diameter from the center of the optical host galaxy. The nearest cluster is NSCS J121800+181254 (with a redshift of 0.37) with an approximate projected linear distance of $\sim$480 kpc from the center of the host galaxy. At $\sim$1.6 Mpc projected linear distance from the center of the optical host galaxy, another cluster WHL J121756.5+181009 (with a spectroscopic redshift of 0.4881; \citep{Ahu20}) is found. The source J1218+1813 has a spectroscopic redshift of 0.139635 \citep{Ahu20}, whereas the two aforementioned galaxy clusters are located at significantly higher redshifts of 0.37 and 0.4881, respectively. This substantial redshift difference suggests that J1218+1813 is unlikely to be a member of these galaxy clusters.
Additionally, J1218+1813 lies at a projected distance of 369 kpc from a compact group of galaxies SDSSCGB 18080 \citep{Mc09}, which has a photometric redshift of $0.24\pm0.05$. This compact group of galaxies is enclosed within a projected distance of 40 kpc, corresponding to an angular scale of 0.17 arcminutes. Given the physical separation of approximately 430 Mpc between J1218+1813 and SDSSCGB 18080 with a redshift difference of $\Delta z = 0.1$, it is unlikely that this compact galaxy group has influenced the circular diffuse emission associated with J1218+1813. The large physical and redshift separation makes any direct environmental interaction between J1218+1813 and SDSSCGB 18080 improbable.

\subsection{Integrated Radio Spectrum}
\label{subsec:spec_map}
The flux densities of J1218+1813 are measured and tabulated in Table \ref{tab:basic}. The results of GMRT and VLA are combined with the results available in the literature to calculate the integrated radio spectrum of J1218+1813. Figure \ref{fig:spec_map2} presents the integrated radio spectrum of J1218+1813. To draw this plot, we used flux densities at various frequencies. We collected 10 measurements of flux densities at 147 MHz (TGSS survey), 402 MHz (GMRT band 3), 647 MHz (GMRT band 4), 888 MHz (ASKAP survey), 1274 MHz (GMRT band 5), 1519 MHz (VLA \textit{L} band), 3000 MHz (VLASS survey), 4160 MHz (VLA \textit{C} band), 6300 MHz (VLA \textit{C} band) and 10000 MHz (VLA \textit{X} band). We fit the broken power law for this integrated radio spectrum (see Figure \ref{fig:spec_map2}). In this spectrum, the spectral energy distribution (SED) of J1218+1813 is predominantly attributed to synchrotron radiation arising from relativistic charged particles spiraling within the magnetic field. The GMRT band 5 image (see in the right panel of Figure \ref{fig:GMRT_band_3_5}), VLASS image, VLA at the \textit{C} band (see the upper left panel of Figure \ref{fig:VLA_GMRT}) and VLA at the \textit{X} band (see the lower left panel of Figure \ref{fig:VLA_GMRT}), resolves the compact radio core component of J1218+1813. We measured the flux density of the resolved core radio component (tabulated in Table \ref{tab:basic}) with task \textit{JMFIT} at 1274 MHz (GMRT band 5 image), VLASS at 3000 MHz, and VLA at 6000 MHz and 10,000 MHz. The integrated radio spectrum (or power-law spectra) of the resolved radio core component between 1274 MHz and 10,000 MHz gives the best-fit spectral index of $0.80\pm0.07$ (see Figure \ref{fig:spec_map3}). Throughout the current paper, we follow $S_{\nu}$ $\propto$ ${\nu}^{-\alpha}$ convention for spectral index measurement.

\subsection{Spectral Index Map}
\label{subsec:spec_map2}
\begin{figure}
\vbox{
\centerline{
\includegraphics[width=9.5cm, origin=c]{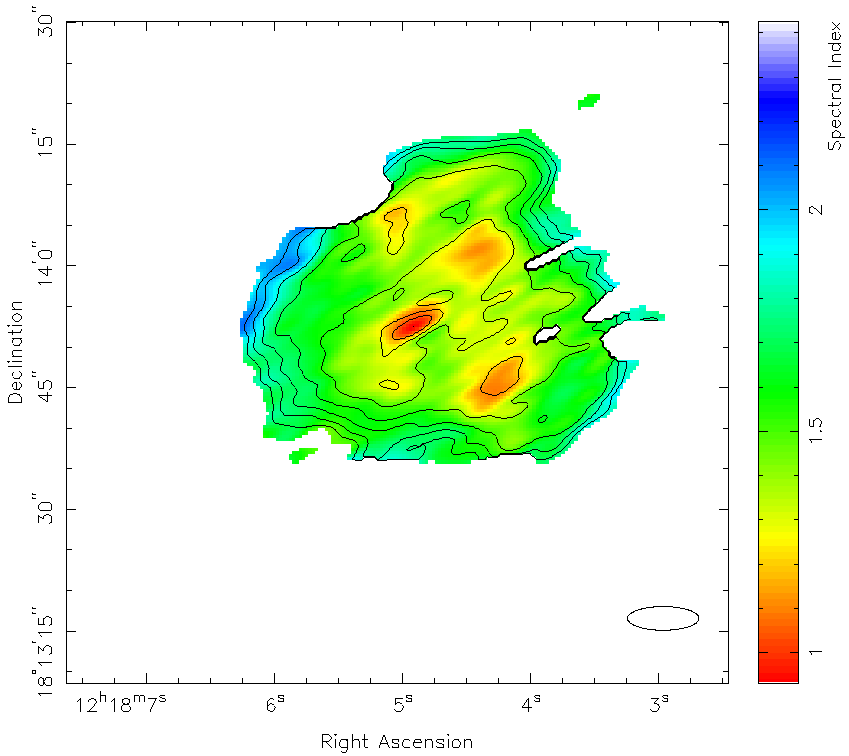}
}
}
\caption{Spectral index map of J1218+1813 between 1274 and 10,000 MHz with an angular resolution of 8.8$'' \times 2.92''$. The contour map at 6000 MHz is overlayed with the spectral index map.}
\label{fig:spec_map}
\end{figure}

To study the spectral nature of the diffuse emission of J1218+1813, we create a spectral index map (see Figure \ref{fig:spec_map}) of the source by using 1274 MHz (GMRT band 5), 6000 MHz (VLA \textit{C} band), and 10,000 MHz (VLA \textit{X} band) with a common resolution of 8.8$'' \times 2.92''$. We used Broadband Radio Astronomy Tools software \citep{harwood_2013MNRAS}\footnote{http://www.askanastronomer.co.uk/brats/} to create this spectral index map.
The contour plot in Figure \ref{fig:spec_map} shows the VLA \textit{C} band image of J1218+1813 at 6000 MHz.
A 3$\sigma$ clipping threshold was applied to ensure the reliability and accuracy of the resulting map. For lower resolution and poor sensitivity at other frequencies images (e.g., 147 MHz image, 888 MHz ASKAP image, or 1519 MHz VLA \textit{L} band image), we did not use them for creating the spectral index map. 
The spectral index of J1218+1813 steepens as we go along the circular periphery structure of J1218+1813. The spectral index map in Figure \ref{fig:spec_map} shows the spectral index variation from 1.1 to 1.6 for J1218+1813 along the inner structure. The spectral index varies from 1.6 to 1.85 around the circular periphery or other than the inner structure of the source J1218+1813.
Similar ranges of spectral indices were also observed along diffuse emission for two circular diffuse sources (J1407+0453 and J1507+3013) identified by \citet{Ku23a, Ku23b}.  
The core region in the spectral index map is seen as steep ($\sim1$; see Figure \ref{fig:spec_map}). Here, the core region may be mixed with nearby radio emission, resulting in the spectral index of the core being steeper. However, we separately measured the spectral index of the core (discussed in sec \ref{subsec:spec_map}), which is also steep ($\alpha = 0.80\pm0.07$).

\begin{figure}
\vbox{
\centerline{
\includegraphics[width=9.0cm, origin=c]{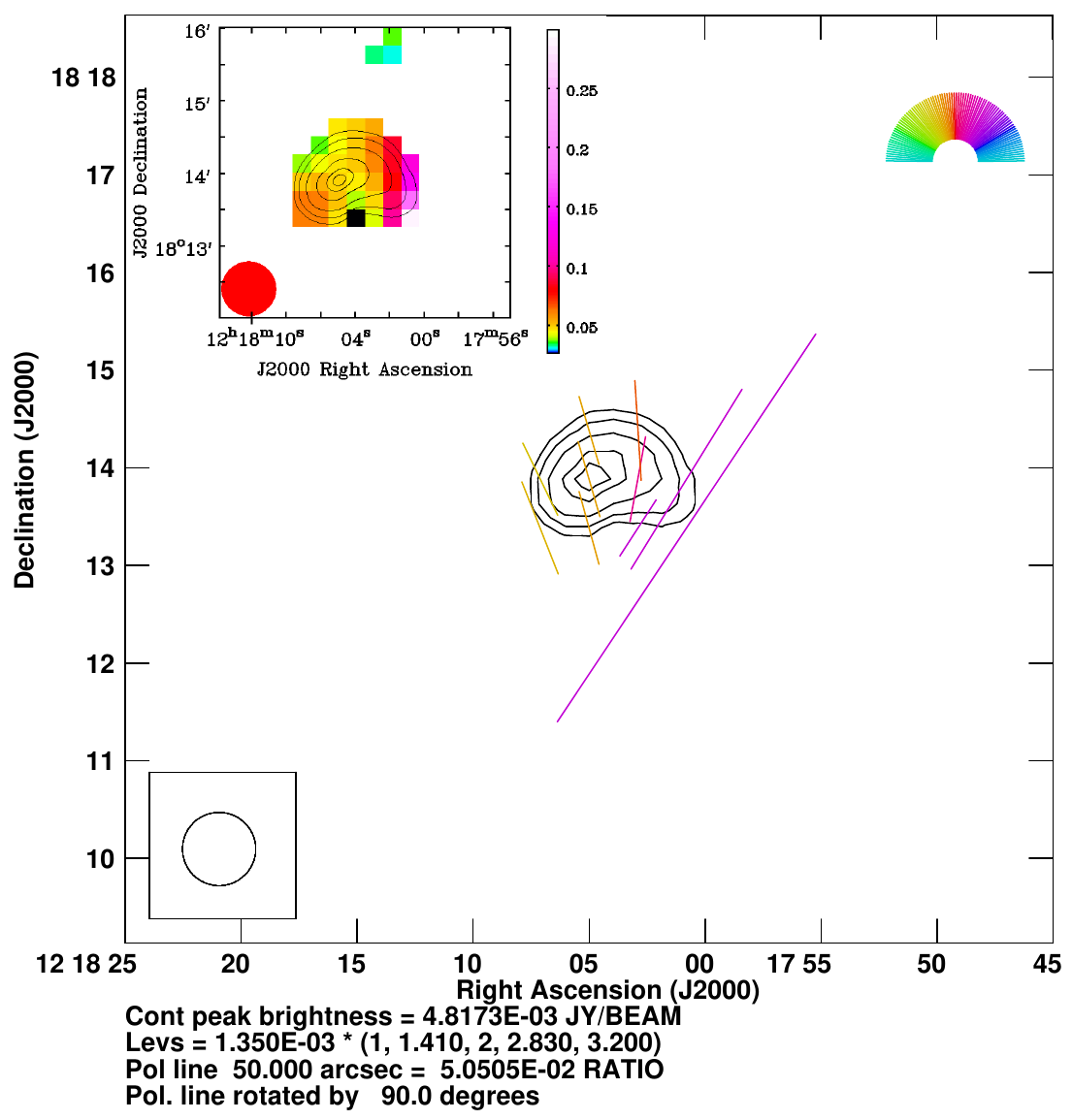}
}
}
\caption{Polarized flux map in black contour overplotted with polarization line vectors (rotated by 90$^{\circ}$). The contour levels and contour peak brightnesses are shown below in the image. The inset image shows the polarized flux map in black contour overplotted with fractional polarization in color scale. The contour levels are at 3$\sigma$ $\times$(1, 1.41, 2, 2.83, 3.20, 4.00), where $\sigma=0.45$ mJy beam$^{-1}$.}
\label{fig:polarization}
\end{figure}

\begin{table}
\centering
\renewcommand{\arraystretch}{1.3} 

\caption{Radio properties of J1218+1813} 
\label{tab:basic} 
\begin{tabular}{cc}
\hline
Properties & Values \\ 
\hline
LAS, Total physical extent & 70$''$, 180 kpc \\  
$S_{\rm 147 MHz}$ (mJy) & 1058 $\pm$ 53 \\ 
$S_{\rm 402 MHz}$ (mJy) & 530 $\pm$ 27 \\ 
$S_{\rm 647 MHz}$ (mJy) & 398 $\pm$ 19 \\
$S_{\rm 888 MHz}$ (mJy) & 278 $\pm$ 14 \\ 
$S_{\rm 1274 MHz}$ (mJy) & 182 $\pm$ 9 \\ 
$S_{\rm 1519 MHz}$ (mJy) & 157 $\pm$ 16 \\
$S_{\rm 3000 MHz}$ (mJy) & 54 $\pm$ 4 \\ 
$S_{\rm 4160 MHz}$ (mJy) & 29 $\pm$ 2.7 \\ 
$S_{\rm 6300 MHz}$ (mJy) & 13.3 $\pm$ 1.8 \\
$S_{\rm 10000 MHz}$ (mJy) & 5.5 $\pm$ 1.05 \\
$S_{\rm 1274 MHz}^{Core}$ (mJy) & 2.7 $\pm$ 0.8 \\ 
$S_{\rm 3000 MHz}^{Core}$ (mJy) & 1.1 $\pm$ 0.09 \\ 
$S_{\rm 6000 MHz}^{Core}$ (mJy) & 0.55 $\pm$ 0.06 \\
$S_{\rm 10000 MHz}^{Core}$ (mJy) & 0.41 $\pm$ 0.015 \\
$P_{\rm 1274 MHz}^{\rm Total}$ ($\times$ $10^{24}$ W Hz$^{-1}$) & 9.6 $\pm$ 0.4 \\ 

\hline
\end{tabular}
\\

\textbf{Note}. Angular size refers to the largest angular size (LAS) or angular diameter, as measured from the 3$\sigma$ contours of the VLA FIRST image at 1400 MHz.
\end{table}

The break frequency is calculated by fitting the fluxes at 147 MHz (TGSS survey), 402 MHz (GMRT band 3), 647 MHz (GMRT band 4), 888 MHz (ASKAP survey), 1274 MHz (GMRT band 5), 1519 MHz (VLA \textit{L} band), 3000 MHz (VLASS survey), 4160 MHz (VLA \textit{C} band), 6300 MHz (VLA \textit{C} band) and 10,000 MHz (VLA \textit{X} band) (see Figure \ref{fig:spec_map2}). The flux values at various frequencies are tabulated in Table \ref{tab:basic}. The break frequency shown in Figure \ref{fig:spec_map2} is approximately 1.56 GHz. 

\begin{figure*}
\vbox{
\centerline{
\includegraphics[width=7.9cm, origin=c]{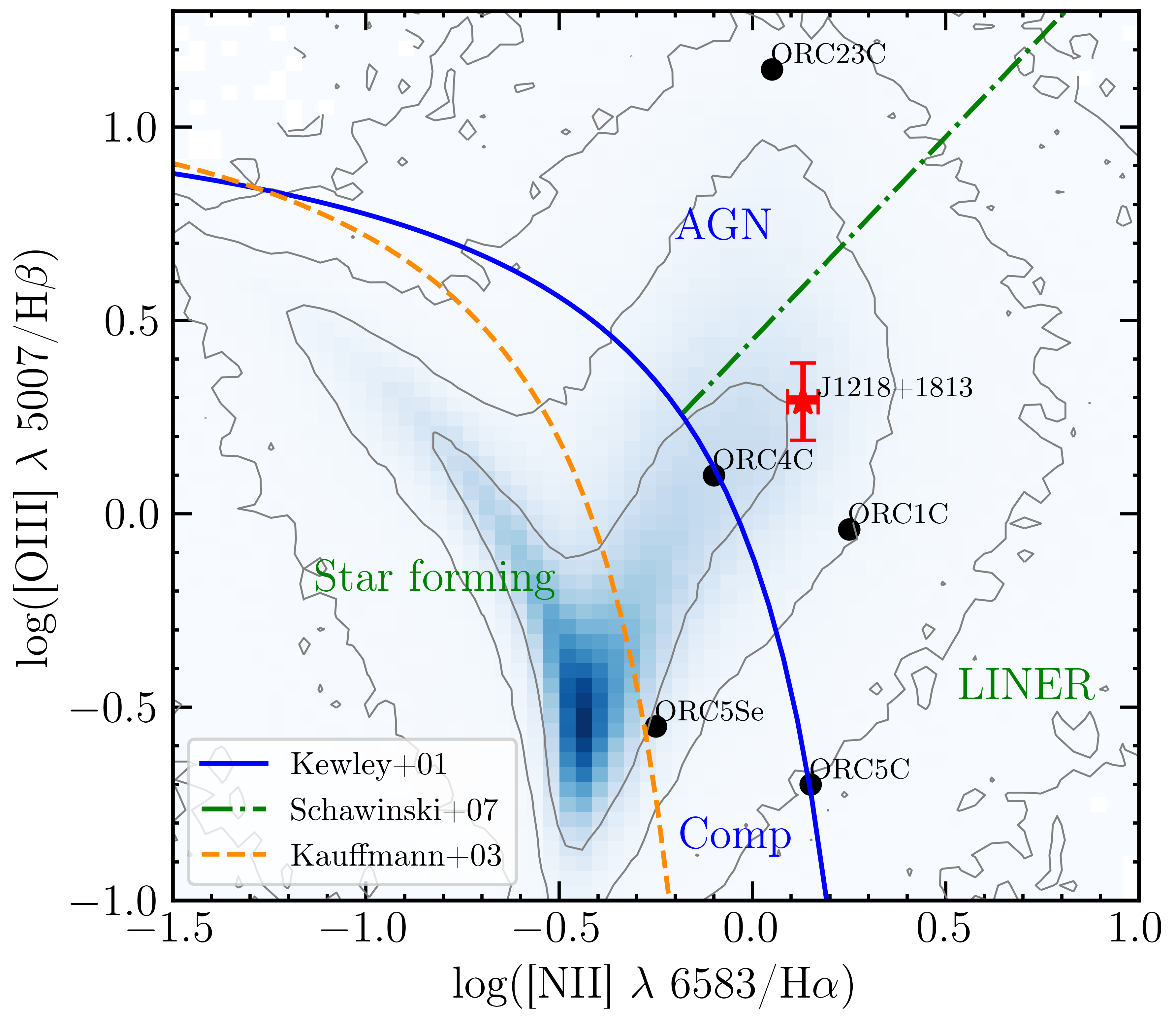}
\includegraphics[width=11cm, origin=c]{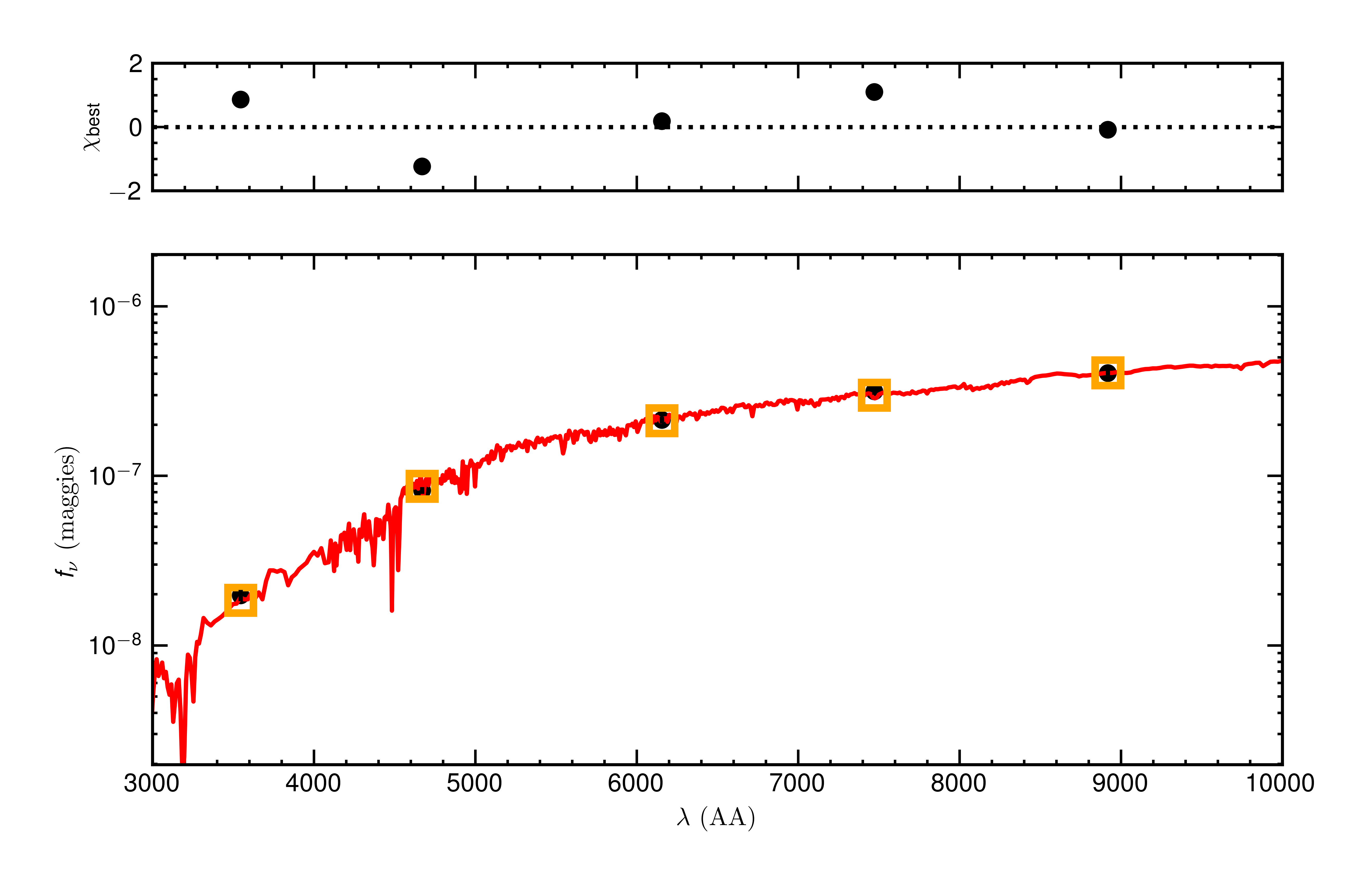}
}
}
\caption{Left: The BPT diagram of previously detected ORC host galaxies along with the source (J1218+1813) presented in the current paper. The red color point represents J1218+1813. Lower right: Spectrophotometric SED of J1218+1813 using the spectrum and $ugriz$ photometry from SDSS. Top right: The normalized residual (or the chi-like statistic) in the photometric fitting.}
\label{fig:BPT}
\end{figure*}
\subsection{Radio Luminosity Measurements}
\label{subsec:luminosity}
The radio luminosity is calculated using the standard formula \citep{Do09}
\begin{equation}	
    L_{\textrm{rad}}=4\pi{D_{L}}^{2}S_{1400}(1+z)^{-(\alpha+1)}
\end{equation}
where $z$ is the redshift of the radio galaxy, $\alpha$ is the spectral index ($S \propto \nu^{-\alpha}$), $D_{L}$ is luminosity distance to the source in meters (m), and $S_{1274}$ is the flux density (W m$^{-2}$ Hz$^{-1}$) at 1274 MHz. Here, we took the average spectral index of $1.5\pm0.1$ from the spectral index map of J1218+1813 (presented in Figure \ref{fig:spec_map}). The measured radio luminosity of J1218+1813 at 1274 MHz is $9.6\pm0.4\times10^{24}$ W Hz$^{-1}$, suggesting that it has a luminous AGN. The radio luminosity of J1218+1813 is similar to the recently discovered circular-symmetry source J1507+3013 \citep{Ku23a} and the horseshoe-shaped diffuse source by \citet{Ku23b}. 

\subsection{Polarization Study}
\label{subsec:polarisation}
Since high-resolution polarization data for J1218+1813 were unavailable, we utilized the NVSS image at 1400 MHz for the polarization study of the source. From the Stokes $Q$ and  $U$ images of the NVSS, the polarization intensity and polarization angle maps of J1218+1813 are derived as follows:
\begin{eqnarray}
&& P=\sqrt{U^2+Q^2} \nonumber\\
&& \Psi = \frac{1}{2} tan^{-1}{\left(\frac{U}{Q}\right)}
\label{eq:UQ}
\end{eqnarray}
Although the NVSS radio image exhibits nearly uniform sensitivity, certain regions may experience reduced sensitivity due to dynamic range limitations. To mitigate this, we applied a $3\sigma$ blanking threshold to the total intensity image of J1218+1813 and subsequently used this blanked image as a mask for the polarization intensity map. The resulting polarization intensity image extends beyond a single-beam area, ensuring that the analysis does not focus solely on the brightest part of the low-brightness source, which could otherwise introduce a bias toward higher fractional polarization values ($F_P$). Finally, we calculate the mean total intensity flux and the mean polarized intensity flux for J1218+1813, from which the fractional polarization was determined as $F_P = P/I$. This gives a fractional polarization of 0.05 (5\%) with the polarized flux ($P$) of 8.64 mJy and total flux ($I$) of 173 mJy in the NVSS map at 1400 MHz. As shown in the inset image of Figure \ref{fig:polarization}, the fractional polarization value of J1218+1813 predominantly ranges between 0.04 and 0.06, represented by the yellow color in the polarization line vector plot of Figure \ref{fig:polarization}. In contrast, the fractional polarization increases from 0.08 to 0.28, as indicated by the purple color in the same plot. A polarization line vector with a full 50$''$ length corresponds to a region with fractional polarization of approximately $\sim$5\% (see Figure \ref{fig:polarization}), while shorter vectors indicate lower fractional polarization values, and vice-versa.

\subsection{Analysis of Optical Properties and Spectral Energy Distribution (SED) Modeling}
\label{subsec:SED}

To examine the properties of the optical host galaxy of J1218+1813, we look for optical spectral and SED modeling. We first acquired the spectrum and $ugriz$ photometry data from the SDSS server. After subtracting the stellar continuum, we estimated the emission-line properties by fitting Gaussian profiles to spectral lines using the Levenberg-Marquardt method \citep{lm63}.
To measure stellar velocity dispersion, we modeled the spectra using the penalized PiXel-Fitting method (pPXF; \citealt{cap04,cap17}). We estimated the stellar velocity dispersion ($\sigma_{\ast}$) as $211\pm9$ $\rm km~s^{-1}$.

To estimate the stellar mass and track the star formation history, we employed spectrophotometric SED modeling of J1218+1813 with the flexible stellar population synthesis (FSPS; \citealt{con09,con10}) library using the SDSS spectrum and $ugriz$ photometry with known redshift (see the right panel of Figure \ref{fig:BPT}). The top right panel indicates the normalized residual (or the chi-like statistic) in the photometric fitting. We used the MILES spectral library \citep{miles}, the MIST isochrone \citep{choi16}, and \citet{dl07} silicate-graphite-PAH grain model as a dust emission model with FSPS. Bayesian posteriors were estimated and sampled using the dynamic nested sampling tool, \textit{dynesty} \citep{sp20}. In this study, we adopted the initial mass function from \citet{k01}. To describe the star formation history, we utilized a delayed $\tau$-model in which the star formation rate (SFR) decreases as $t_\mathrm{age}\,e^{-t_\mathrm{age}/\tau}$, where $t_\mathrm{age}$ represents the age of the galaxy. We apply a log-uniform prior on $t_\mathrm{age}$ to ensure that the galaxy is not younger than 100 million years when observed and does not form before the beginning of the Universe ($0.1~ \mathrm{Gyr} < t_\mathrm{age} < t_\mathrm{Universe}(z)$, where $t_\mathrm{Universe}(z)$ is the age of the Universe at the redshift ($z$) of the galaxy).

We tabulated the best-fit parameters with 1$\sigma$ error in Table~\ref{tab:optical}. We estimated the current SFR and SFR 1~Gyr ago using the following equations:
\begin{eqnarray}
    \mathrm{SFR} &\,=\,& \mathrm{SFR_{max}}\frac{t_\mathrm{age}}{\tau}(e^{1-t_\mathrm{age}/\tau}) \\
  \mathrm{SFR_{max}} &\,=\,& M_*(e^{-1}\times10^{-9}~M_\odot {\rm yr}^{-1})\;\times \nonumber \\
  ~ & ~ & \tau^{-1}\left[1-\left(1+\frac{t_\mathrm{age}}{\tau}\right)e^{-t_\mathrm{age}/\tau}\right]^{-1}
\end{eqnarray}
where $t_\mathrm{age}$ and $\tau$ are in Gyr and \textit{M}$_*$ is in $\textit{M}_\odot$.\\

We also estimated the excitation index (EI) to classify the host of J1218+1813 as a high-excitation radio galaxy (HERG) or low-excitation radio galaxy (LERG) using the following equation,
 \begin{equation}
 \rm EI=Log\frac{[OIII]}{H_{\beta}}-\frac{1}{3}Log\frac{[NII]}{H_{\alpha}}+ Log\frac{[SII]}{H_{\alpha}}+Log\frac{[OI]}{H_{\alpha}}
 \end{equation}
EI quantifies the relative strength of high- and low-excitation optical emission lines in the spectrum of a galaxy. An EI value less than 0.95 ($\mathrm{EI} < 0.95$) typically signifies a LERG, characterized by weak or absent high-excitation lines and radiatively inefficient accretion. In contrast, sources with EI greater than 0.95 ($\mathrm{EI} > 0.95$) are classified as HERGs, exhibiting strong high-excitation lines indicative of radiatively efficient accretion processes. The measured EI for the central optical galaxy of J1218+1813 is 2.62. So, based on EI, J1218+1813 is a HERG source. We also present the Baldwin–Phillips–Terlevich (BPT) diagram of J1218+1813 in the left panel of Figure~\ref{fig:BPT}, which provides a diagnostic tool to distinguish between different ionization mechanisms based on optical emission-line ratios. In this diagram, we have additionally included a sample of previously detected ORCs for comparative analysis. The positioning of J1218+1813 relative to the ORCs population helps to assess its ionization state and the dominant source of excitation, such as star formation, AGN activity, or a composite nature. The red color represents the source, J1218+1813, presented in the current paper. According to this diagram, J1218+1813 falls in a low-ionization nuclear emission-line region (LINER) similar to one of the ORC (ORC J2103--6200; ORC1).

\begin{figure*}
\vbox{
\centerline{
\includegraphics[width=16.0cm, origin=c]{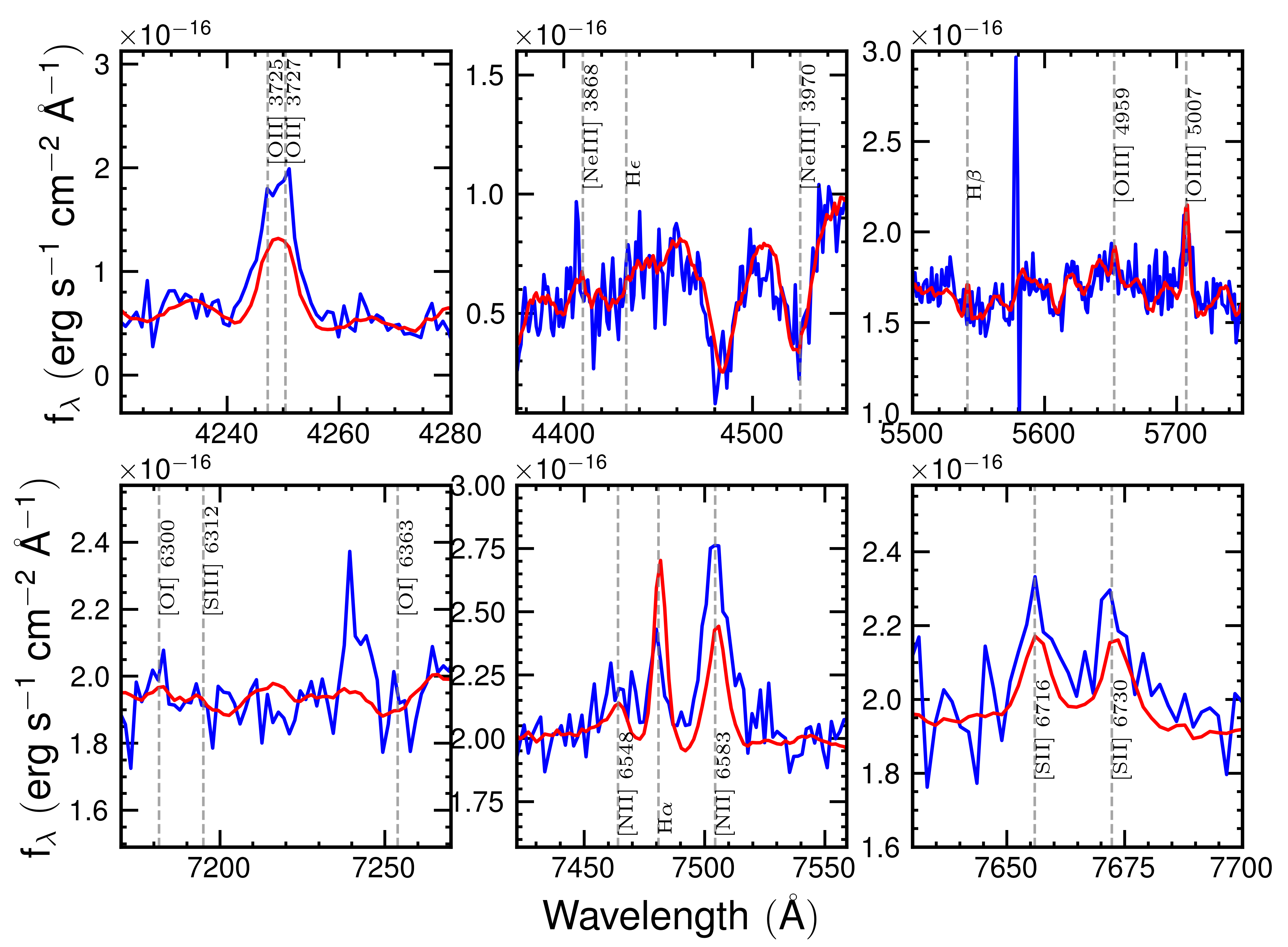}
}
}
\caption{Optical spectra of the host of J1218+1813, with best-fitted detected emission lines. The blue lines depict the observed spectra and the red lines represent the best-fitted spectra.}
\label{fig:spec}
\end{figure*}

We also estimated the luminosities of the emission lines H$_{\alpha}$, H$_{\beta}$, and [OII] using the fitted flux integrated into the line (see Figure \ref{fig:spec}) obtained from the fitting of the SDSS spectra. Since at least some of the H$_{\alpha}$ emissions arise from processes other than photoionization from young stars, we estimated the upper limit of the SFR using the H$_{\alpha}$ and [OII] line luminosities, as tabulated in Table \ref{tab:optical}.

We estimated the black hole mass ($\rm \textit{M}_{BH}$) of the host galaxy of J1218+1813 using the following  $\rm \textit{M}_{BH}$--$\sigma_{\ast}$ relation \citep{Mcc13}:

\begin{equation}
	\rm\log\left(\frac{\textit{M}_{\text{BH}}}{\textit{M}_{\odot}}\right)=(8.32\pm 0.05) +(5.64 \pm 0.32)~log\left(\frac{\sigma_{\ast}}{200~\text{km~s}^{-1}}\right)	
\end{equation}
where $\sigma_{\ast}$ is the velocity dispersion, and \textit{M}$_{\odot}$ is the stellar mass ($\sim$2 $\times$$10^{31}$ kg). Using the estimated stellar velocity dispersion, the calculated black hole mass of the host galaxy of J1218+1813 is $2.8 \pm 0.8 \times 10^{8}$ \textit{M}${_\odot}$. A slightly higher black hole mass (5.8 $\times10^8$ \textit{M}$_{\odot}$) is measured for J1407+0453, a source with a horseshoe-shaped inner ring of diffuse emission \citep{Ku23b}. 

\subsection{Infrared Properties}
\label{subsec:IR}
We used the Wide-field Infrared Survey Explorer \cite[WISE;][]{Wr10, Cu21} catalog to study the infrared properties of the host galaxy. The WISE uses the W1, W2, W3, and W4 bands at wavelengths of 3.4 $\mu$m, 4.6 $\mu$m, 12 $\mu$m, and 22 $\mu$m, respectively. The host galaxy has u -- r = 3.24, W1 -- W2 = 0.153 and W2 -- W3 = 1.8. Using infrared colors, galaxies are usually classified as LERG if W1 -- W2 $<$ 0.5 and 0.2 $<$ W2 -- W3 $<$ 4.5 or HERG if --0.2 $<$ W1 -- W2 $<$ 1.55 and 1.32 $<$ W2 -- W3 $<$ 4.26 \citep{Pr18}. Based on infrared colors, it is uncertain whether J1218+1813 is a LERG or HERG because it falls into both classes.

\begin{table}
\centering
\renewcommand{\arraystretch}{1.3} 
\caption{Optical Spectral Properties of J1218+1813.} 
\label{tab:optical} 
\begin{tabular}{cc}
\hline
Properties & Values \\ 
\hline
Redshift ($z$) & 0.139635   \\  
BPT classification& LINER\\
$\rm F_{OII}(10^{-17} erg/s/cm^{2})$& $38.14\pm6.69$\\
$\rm F_{OIII,\lambda 5007}(10^{-17} erg/s/cm^{2})$& $30.17\pm3.54$\\
$\rm F_{H\beta}(10^{-17} erg/s/cm^{2})$& $15.44\pm3.21$\\
$\rm F_{H\alpha}(10^{-17} erg/s/cm^{2})$& $54.40\pm4.24$\\
$\rm F_{NII,\lambda 6583}(10^{-17} erg/s/cm^{2})$& $73.64\pm3.75$\\
$\sigma_{\ast} (\rm km~s^{-1})$& $211\pm9$ \\
$\rm\log(M_{\text{BH}}/M_{\odot})$ & $8.45\pm0.12$\\
$\rm L_{H_{\alpha}} (erg~s^{-1})$& $(2.83\pm0.22)\times10^{40}$\\
$\rm L_{[OII]} (erg~s^{-1})$& $(1.99\pm0.35)\times10^{40}$\\
$\rm SFR_{H_{\alpha}} (M_{\odot}~yr^{-1})$& $<0.224^{0.017}$ \\
$\rm SFR_{[OII]} (M_{\odot}~yr^{-1})$& $<0.28^{0.05}$\\
\hline
&\hspace{-5cm}Best-Fitted Stellar Properties from SED\\ 
\hline
Properties & Values \\ 
\hline
$\rm\log(M_{\text{*}}/M_{\odot})$ & $11.49\pm0.011$ \\ 
Stellar metallicity, logzsol& --0.44\\
Diffuse dust V-band optical depth& 0.26\\
$\rm t_{age} (Gyr)$& 8.58\\ 
$\tau$& 0.11\\
SFR$\rm_{now} (M_{\odot}~yr^{-1})$& 0\\
SFR$\rm_{1 Gyr} (M_{\odot}~yr^{-1})$& 0\\
$e_bv_sfd$& 0.0353\\
\hline
\end{tabular}
\end{table}

%%%%%%%%%%%%%%%%%%%%%%%%%%%%%%%%%%%%%%%%%%
\section{Discussion}
\label{sec:discussion}
In this section, we discuss the unique characteristics of J1218+1813. Some possible formation scenarios for the reported source J1218+1813 in the current paper are also discussed below. 

\subsection{Wing-like Emission in J1218+1813}
\label{sec:back}
As discussed in Sections \ref{sec:result_VLA} and \ref{sec:result_GMRT}, J1218+1813 has a wing-like structure or SW emission. In the superimposed image of J1218+1813, as shown in the lower right panel of Figure \ref{fig:VLA_GMRT}, combining VLA \textit{C} band data ($\sim$6000 MHz) in color scale with GMRT Band 4 in contour at $\sim$402 MHz, the emission from the northern and southern SWs are clearly visible. The secondary wing emissions are typically seen in X-shaped radio galaxies and various models have been proposed for the origination of secondary wings in X-shaped radio galaxies, viz., the backflow model \citep{Le84, Ca02}, the twin AGN model \citep{Ba80, Wo95}, slow jet reorientation (precession) model \cite{Ho20}, the rapid jet reorientation model \citep{Me02}, etc. For J1218+1813, the plasma may gain backflow after hitting the surrounding denser environment and spreading towards the minor axis direction, resulting in the formation of a wing feature. It is important to note that Fanaroff–Riley type II (FRII) sources are most commonly associated with winged or X-shaped radio galaxies \citep{Ca02, Sa09, Go12, Co20, bh22, Sh24}, which are well explained by the backflow model. However, recent studies have identified a few Fanaroff–Riley type I (FRI) sources that also exhibit X-shaped structures \citep{Sa09, Go12, Co20}, suggesting that the formation mechanisms of such morphologies may not be exclusively related to FRII sources. Seeing the structure of J1218+1813, it is very complex to classify the source as FRI or FRII.

\subsection{Restarted or Recurrent Activity in J1218+1813?}
\label{sec:restarted}
The core of J1218+1813 has a steep spectral index ($\alpha =0.8$) (discussed in Section \ref{subsec:spec_map}). 
The steep spectral index of the core of J1218+1813 can also be seen in the spectral index maps of J1218+1813 as discussed in Section \ref{subsec:spec_map2}. The physical reasons behind the flat and steep spectrum cores in AGN are primarily related to synchrotron emission and absorption processes, respectively. In many radio galaxies with powerful AGNs, the core often exhibits a flat or inverted spectrum ($\alpha <0.5$) due to synchrotron self-absorption \citep{Bl79, Ke81, Od98, Sh24}. This occurs when the emitted synchrotron radiation is reabsorbed by the same population of relativistic electrons, preventing lower-frequency radiation from escaping. As a result, the core appears compact and dense, with a high optical depth at lower frequencies, leading to a flattened observed spectrum. On the other hand, in lower-power AGN, the core can have a steeper spectrum (e.g., 0.8 or steeper) if optically thin synchrotron emission dominates. In this case, the radiation is not significantly absorbed, allowing us to observe emission from an extended jet region rather than just the self-absorbed core. The steep spectrum of the core of J1218+1813 may be dominated by extended features (such as unresolved jets or lobes near the core). This suggests that the AGN of J1218+1813 may be experiencing restarted or recurrent activity. 

J1218+1813 may be a relic of past AGN activity, i.e., the remnant emission of a once-active supermassive black hole (SMBH) that has since switched off or entered a quiescent phase. AGN jets inject relativistic electrons into the surrounding medium. When the AGN ceases activity, the lobes gradually expand and fade as the electrons lose energy via synchrotron and inverse Compton cooling.
The steep spectral index observed in J1218+1813 ($\alpha = 1.1–1.8$, along the outer structure) supports this scenario because older, radiatively aged plasma typically exhibits steep spectra. There is a lack of a clear jet structure in J1218+1813, suggesting that the AGN has long been inactive and the core shows signs of a restarted AGN or episodic activity (showing the inner feature with a lower spectral index than the outer). Sources such as J1835+620 and J1453+3308, known for their restarted AGN activity, display reduced fractional polarization ($\sim 0.03-0.07$) near the inner lobes, comparable to the values observed in J1218+1813. This similarity suggests the presence of mixed plasma populations with aged electron components \citep{Ko12}.

\subsection{Possible Formation Scenarios for J1218+1813}
\subsubsection{Is J1218+1813 a Result of AGN Jet Precession?}
\label{sec:S-shaped}

The circularly symmetric diffuse structure of J1218+1813 may be attributed to AGN jet precession. Precessing jets can deposit plasma over a broader region, forming a cocoon-like structure with enhanced lateral extent rather than propagating in a specific direction \citep{Ha01}. Jets with larger precession angles are known to spread out more significantly, creating such wide and diffuse regions. Conversely, jets with shorter precession periods are typically more centrally concentrated because they fail to sustain momentum long enough to overcome the resistance posed by the surrounding IGM \citep{Br78, Ca02, La06}. 

In the case of J1218+1813, the relatively small angular size of 70 arcsec suggests that the jets may have been unable to extend far from the core due to interactions with dense IGM \citep{Ha06}. This compact plasma distribution near the center indicates a possible correlation with a shorter jet precession period. Such a scenario aligns with theoretical models predicting that reduced precession periods limit jet propagation and promote localized diffusion of the synchrotron-emitting plasma \citep{Ho20, No23}.

A recent simulation study by \citet{No23} indicates that a precessing jet may take approximately $t \sim 100$ Myr to turn off along a line of sight parallel to the precession axis. Shortly after the jet's activity ceases, a complete ring-like emission structure may appear. During this phase, the source exhibits a spectral index of $\alpha \sim 0.8$. As the plasma ages, the ring-like structure evolves without significant expansion and its spectral index steepens to values within the range $0.8 < \alpha < 1.2$, consistent with synchrotron aging in a region where the jets no longer inject momentum.

For J1218+1813, the observed spectral index within its diffuse circular structure lies in the steep range $1.1 < \alpha < 1.6$ (see Figure \ref{fig:spec_map}), consistent with aged synchrotron plasma. Additionally, the core of J1218+1813 exhibits a spectral index of $\alpha = 0.80$ (see Figure \ref{fig:spec_map3}). These spectral characteristics, plasma distributions, and spectral properties collectively support the scenario that J1218+1813 may indeed be influenced by AGN jet precession, reinforcing the role of AGN feedback in shaping its circularly diffuse radio emission \citep{No23, Ko21}.

\subsubsection{Is J1218+1813 a Result of a Giant Blast Wave?} \label{sec:blast_wave}
As discussed in Section \ref{subsec:optical}, the optical galaxy hosting J1218+1813 is located near the radio core (see Figure \ref{fig:optical}). The circular and diffuse structure of J1218+1813 may originate from synchrotron emission produced by accelerated electrons from a shock induced by the central galaxy \citep{Do02, Ca12, Ri21, Ko21, Ya24}. Such a spherical shock could result from a binary SMBH merger in the central host galaxy \citep{Bo12}. Recent simulations \citep{Do23} also suggest that such shock-driven circular diffuse emission can account for structures such as J1218+1813.

The optical properties of J1218+1813 are described in Section \ref{subsec:SED} and summarized in Table \ref{tab:optical}. J1218+1813 exhibits a strong [O II] emission line with an [O II] flux of $38.14\pm7$ (in units of 10$^{-17}$ erg s$^{-1}$ cm$^{-2}$ \AA$^{-1}$) and an equivalent width (EW) of $\sim$10 \AA. The stellar velocity dispersion ($\sigma_{\ast}$) is measured as $211\pm9$ km s$^{-1}$ with a stellar mass of $\rm\log(\textit{M}_{\text{*}}/\textit{M}_{\odot})$ = $11.49\pm0.011$. The optical host galaxy of J1218+1813 is estimated to be 8.58 Gyr old. A spectroscopic study by \citet{Co24} revealed that the central optical source of one of the ORCs (ORC J1555+2726; ORC4), with a redshift of $z = 0.4512$, emits strong [O II] [3727, 3729 \AA] ionized gas, extending to a radius of 40 kpc with a significant velocity dispersion of 150--200 km s$^{-1}$ and a high EW (50 \AA). The core optical galaxy corresponds to a 1 Gyr old burst of star formation responsible for half of the stellar content of the galaxy. Numerical simulations suggest that this burst could have driven diffuse radio emission through a forward shock created by a strong wind \citep{Co24}.
For J1218+1813, the current star SFR (SFR$\rm_{now}$) and the SFR at 1 Gyr ago (SFR$\rm_{1 Gyr}$) are estimated, showing no evidence of recent star formation or a burst in the galaxy. The SFR in the core optical galaxy of J1218+1813, derived from H$\alpha$ and [O II], is also very low ($<$ 0.2; see Table \ref{tab:optical}). A detailed optical investigation of J1218+1813 is required to better understand the ionized gas emission surrounding the optical host galaxy.

Although J1218+1813 is not embedded in a dense cluster environment, local turbulence induced by galaxy interactions or a merger-driven shock could still disrupt the magnetic field, lowering the fractional polarization while preserving detectable polarized flux \citep{Burn1966, Murgia2001, Owen2014} as we can see for J1218+1813.
The polarization line vectors generally align with the magnetic field direction in aged radio lobes, reflecting the underlying magnetic field geometry \citep{Laing1988}. In Figure \ref{fig:polarization}, the vectors in yellow and red regions reveal ordered magnetic field structures, whereas those in purple regions exhibit greater directional variation, coinciding with higher fractional polarization values ($\sim$10--20\%). This shift may result from internal magnetic complexity, enhanced turbulence, or reacceleration processes within J1218+1813 itself \citep{Ensslin1998, Hardcastle2013}.

A deeper polarization study incorporating Faraday rotation measure (RM) mapping with high-resolution imaging could provide crucial insights into the intervening magnetic environment and confirm potential interactions with the surrounding medium. Such observations would also help determine whether the observed polarization characteristics are linked to past AGN activity, local environmental effects, or a combination of both.

\subsubsection{Similarities to ORC Formation Mechanisms} \label{sec:ORC_formation} 
Although J1218+1813 does not exhibit the characteristic edge-brightened morphology commonly observed in newly discovered ORCs, several formation scenarios suggest potential similarities between their origins. Various proposed mechanisms for ORC formation may also provide insights into the origin of J1218+1813.

Both J1218+1813 and ORCs may result from large-scale shock waves propagating through the IGM. Such shocks can be triggered by powerful AGN outbursts, merger-induced turbulence, or the collapse of overdense regions, producing diffuse synchrotron emission in circular-symmetry structure \citep{No21a, No21b, No21c, Om22c, Ko21}.

Galaxy interactions and merger events have been suggested as viable triggers for ORC formation, where the resulting merger-driven shocks may re-energize aged relativistic plasma or initiate the expansion of plasma bubbles \citep{Do23, Ya24, Li24}. A similar scenario could explain the circular morphology of J1218+1813, particularly if merger-driven disturbances shaped the synchrotron-emitting plasma.

Episodic AGN outbursts followed by a quiescent phase may contribute to the formation of both J1218+1813 and ORCs. The fading radio lobes of a previously active AGN can evolve into circular, low-surface-brightness features resembling relic radio structures \citep{Ko24a}. Such processes are commonly linked to AGN feedback cycles and energy injection in the surrounding medium.

Furthermore, if J1218+1813 represents a case of relic AGN activity or a merger-driven shock, its study could provide valuable insight into the physical conditions required for ORC formation. Investigating its spectral properties, polarization structure, and surrounding environment may help constrain the broader evolutionary processes responsible for circular radio morphologies. Thus, J1218+1813 may serve as an important analog for understanding the complex physical mechanisms that produce ORCs and other circular-symmetry diffuse synchrotron structures. 

\section{Conclusions}
\label{sec:conclusion}

We present a detailed multiwavelength study of a radio source J1218+1813, which exhibits diffuse, circularly symmetric radio emission surrounding an elliptical galaxy. This peculiar morphology distinguishes it from typical radio galaxies and invites investigation into its origin.

Using deep observations from the uGMRT and VLA, we find that:

\begin{itemize}
    \item The source lacks prominent jet or hotspot structures, and the radio emission appears uniformly distributed in an angular scale of 70$''$ ($\sim$180 kpc).
    
    \item The spectral index analysis indicates a steep spectrum and possible past AGN activity
    
    \item The BPT diagram and optical spectroscopy from SDSS show weak emission lines, placing the host galaxy within the LINER region.
    
    \item We estimate the luminosities of key emission lines (H$\alpha$, H$\beta$, [O\,\textsc{ii}]) and provide an upper limit on the SFR, indicating that star formation is not the dominant energy source.
    \item The lower fractional polarization in J1218+1813 can result from local turbulence induced by galaxy interactions or a merger-driven shock by disrupting the magnetic field. 
    \item The radio morphology and spectral properties suggest possible formation scenarios involving short-period jet precession, giant blast wave, restarted activity, or interaction with the ambient medium.
    
\end{itemize}

Our study highlights J1218+1813 as a rare class of objects in which diffuse emission surrounds a galaxy without any physical association with the cluster.

Wideband radio data using advanced telescopes, such as ASKAP, MeerKAT, GMRT, VLA, and LOFAR, should reveal new samples of sources with diffused emission, improving our understanding of their nature and formation mechanisms.
Further follow-up multiwavelength observations are strongly encouraged to unravel the nature of J1218+1813. In particular, deep X-ray observations will be crucial to identify and study any energetic events within the diffuse source J1218+1813 that may contribute to its extended emission. 

\section*{Acknowledgements}
We thank the anonymous reviewer for helpful suggestions.
We thank the GMRT staff for making the GMRT observations possible. GMRT is run by the National Centre for Radio Astrophysics of the Tata Institute of Fundamental Research. We thank the staff of the National Radio Astronomy Observatory (NRAO) that made Karl G. Jansky Very Large Array (VLA) observations possible. The NRAO is a facility of the National Science Foundation operated under a cooperative agreement by Associated Universities, Inc. S.K. thanks the Department of Science \& Technology, Government of India, for financial support, vide reference number DST/WISE-PhD/PM/2023/3 (G) under the ``WISE Fellowship for PhD" program to carry out this work.

\end{document}